\documentclass{article}
\usepackage{amsfonts}
\usepackage{amssymb}
\usepackage{amsmath}
\usepackage{cite}
\allowdisplaybreaks
\numberwithin{equation}{section}
\DeclareMathOperator{\sgn}{\rm sgn}
\DeclareMathOperator{\diag}{\rm diag}
\DeclareMathOperator*{\res}{\rm res}

\DeclareMathOperator{\Rs}{\mathbb{R}}

\DeclareMathOperator{\Cs}{\mathbb{C}}
\DeclareMathOperator{\Go}{\mathcal{G}}
\DeclareMathOperator{\Lo}{\mathcal{L}}
\DeclareMathOperator{\No}{\mathcal{N}}

\DeclareMathOperator{\Do}{\mathcal{D}}
\DeclareMathOperator{\Vo}{\mathcal{V}}

\DeclareMathOperator{\Ao}{\mathcal{A}}

\DeclareMathOperator{\bk}{\mathbf{k}}
\DeclareMathOperator{\erf}{\rm{erf}}
\def\t#1{\widetilde{#1}}
\def\h#1{\widehat{#1}}
\def\ol#1{\overline{#1}}

\begin{document}

\title{IST of KPII equation for perturbed multisoliton solutions}
\author{M.~Boiti${}^{*}$, F.~Pempinelli${}^{*}$, and A.~K.~Pogrebkov${}^{\dag}$\\
${}^{*}$Sezione INFN, Lecce, Italy\\
${}^{\dag}$Steklov Mathematical Institute, Moscow\\ and National Research University ``Higher School of Economics", Russia}
\date{MSC: 37K10, 37K15, 35C08, 37K40\\
Keywords: Kadomtsev--Petviashvili Equation, Heat Operator,\\
Extended Resolvent, Solitons}
\maketitle

\begin{abstract}
The Direct and the Inverse Scattering Problems for the heat operator with a potential being a perturbation of an arbitrary $N$ soliton potential are formulated. We introduce Jost solutions and spectral data and present their properties. Then, giving the time evolution of the spectral data, the initial value problem of the Kadomtsev-Petviashvili II equation for a solution describing $N$ solitons perturbed by a generic smooth fast decaying potential is linearized.
\end{abstract}

\section{Introduction}

The Kadomtsev--Petviashvili equation in its version called KPII
\begin{equation}
(u_{t}-6uu_{x_{1}}+u_{x_{1}x_{1}x_{1}})_{x_{1}}=-3u_{x_{2}x_{2}},\label{KPII}
\end{equation}
where $u=u(x,t)$, $x=(x_{1},x_{2})$ and subscripts $x_{1}$, $x_{2}$ and $t$ denote partial derivatives, is a (2+1)-dimensional generalization of the celebrated Korteweg--de~Vries (KdV) equation. The KPII equations, originally derived as a model for small-amplitude, long-wavelength, weakly two-dimensional waves in a weakly dispersive medium~\cite{KP1970}, have been
known to be integrable since the beginning of the 1970s~\cite{D1974,ZS1974}, and can be considered as a prototypical (2+1)-dimensional integrable equation. Its integrability results from the existence of the Lax pair
\begin{align}
&\Lo(x,\partial_{x})=-\partial_{x_{2}}+\partial_{x_{1}}^{2}-u(x),\label{heatop}\\
&\Ao(x,\partial_{x})=4\partial_{x_{1}}^{3}-6u\partial_{x_{1}}-3u_{x_{1}}-3\int\limits^{x_{1}}_{\infty}dx'_{1}\,
u_{x_{2}}(x_{1}',x_{2}), \label{230}
\end{align}
so that (\ref{KPII}) is equivalent to  the compatibility condition
\begin{equation}
\Lo_{t}=[\Lo,\Ao]. \label{LaxN}
\end{equation}
Operator (\ref{heatop}) defines the well known equation of heat conduction, or heat equation for short. The corresponding Inverse Scattering Theory (IST) was developed in~\cite{BarYaacov,Lipovsky,Wickerhauser,FS,Grinevich0} in the case of a real potential $u(x)$ rapidly decaying at spatial infinity. This theory is based on the integral equation
\begin{equation}
\Phi (x,\bk)=e^{-i\bk x_1-\bk^{2}x_2}+\int d^{2}y\,\Go_{0}(x-y,\bk)u(y)\Phi (y,\bk), \label{phi0}
\end{equation}
for the Jost solution. Here $\bk\in\Cs$ is the complex spectral parameter and $\Go_{0}(x,\bk)$ is the Green's function of the bare heat operator $-\partial_{x_{2}}+\partial_{x_{1}}^{2}$, given by
\begin{equation}
\Go_{0}(x,\bk) =-\dfrac{\sgn x_{2}}{2\pi }e^{-i\bk x_1-\bk^{2}x_2}\int \!\!ds\,\theta
\bigl(s(s+2\bk^{}_{\Re})x^{}_{2})\bigr)e_{}^{-is(x^{}_1-(s+2\bk)x^{}_2)},\label{G0}
\end{equation}
where $\theta$ is the Heaviside function and integration is along the whole axis. However, class of such potentials is not the most interesting one, since the KPII equation was just proposed in~\cite{KP1970} in order to deal with two dimensional weak transverse perturbation of the soliton solution of the KdV. This solution, as well as multisoliton ones, is known to have а non decaying, ray asymptotic behavior at space infinity (see, e.g., \cite{BPPPr2001a}--\cite{K}). In order to illustrate this behavior, let us notice that equation (\ref{KPII}) is invariant under the Galileo transformation, i.e., if $u(x_1,x_2,t)$ is a solution, then
\begin{equation}
\t{u}(x_{1},x_2,t)=u(x_{1}+\mu x_{2}-3\mu ^{2}t, x_{2}-6\mu t, t), \label{230_1}
\end{equation}
obeys this equation as well for any real constant $\mu$. Then the ray asymptotic behavior follows because any solution $u(x_1,t)$ of the KdV equation solves KPII. This behavior, \cite{BPPPr2001a}, causes the main difficulty for construction of the spectral theory of the operator (\ref{heatop}) with a potential
\begin{equation}
u(x)=u_{N}(x)+u'(x),\qquad x=(x_{1},x_{2}), \label{tu}
\end{equation}
where $u_{N}(x)$ is some $N$ soliton potential and $u'(x)$ is a smooth and rapidly enough decaying function of its variables, that can be considered a perturbation of the soliton potential. Indeed, let us consider the case of the pure one soliton potential, $u(x)=u_1(x_1)$. Then the Jost solution has the form $\Phi(x,\bk)=e^{i\bk x_1+\bk^{2}x_2}\chi(x_1,\bk)$, where $\chi(x_1,\bk)$ is a function of only one spatial variable. Inserting this function into (\ref{phi0}) we see that the integral with respect to $y_2$, as follows from (\ref{G0}), is divergent. Therefore, already for the simplest case of the potential of the type (\ref{tu}) the standard integral equation (\ref{phi0}) for Jost solutions is ill defined, as its kernel does not exist.

Our aim here is the construction of a spectral theory of the KPII equation that also includes solitons, as already successfully done for the KPI equation~\cite{BPP2006b}. Following the framework of that article we develop a generalization of the standard IST: the scattering on a nontrivial background. More precisely, we  take, from the literature, the known generic expressions of the pure $N$ soliton potential $u_N(x)$ and the corresponding Jost solution $\Phi_{N}(x,\bk)$ of the operator $\Lo_{N}$, i.e., $\Lo$ in (\ref{heatop})  with $u=u_N$. Such potentials and Jost solutions can be constructed either by means of rational similarity transformations of the spectral data, or by the $\tau$-function approach, or by twisting transformations (see~\cite{BPPPr2001a,BK,BPPP2009} and Sec.~\ref{NpJ} here), as all these procedures lead to the same result, as it was shown in~\cite{equivKPII}. Then we define the Jost solution of the operator $\Lo$ with potential $u$ (\ref{tu}) by means of the following integral equation
\begin{equation}
\Phi (x,\bk)=\Phi_{N}(x,\bk)+\int dy\,\Go_{N}(x,y,\bk)u'(y)\Phi (y,\bk), \label{tphi}
\end{equation}
where $\Go _{N}(x,y,\bk)$ is the Green's function of the operator $\Lo_{N}$, i.e., solution of the differential equation
\begin{equation}
\bigl(-\partial_{x_{2}}^{}+\partial_{x_{1}}^{2}-u_{N}(x)\bigr)\Go_{N}(x,x',\bk)=\delta (x-x'), \label{green}
\end{equation}
and its dual, fixed by the condition that
\begin{equation}
G_{N}(x,x',\bk)=e_{}^{i\bk(x_{1}-x_{1}')+\bk^{2}(x_{2}-x_{2}')}\Go_{N}(x,x',\bk) \label{Green}
\end{equation}
is bounded with respect to the variables $x,x'\in\Rs^{2}$ and $\bk\in\Cs$ and has finite limits at infinity (``boundedness condition"). In contrast to (\ref{phi0}) the integral equation (\ref{tphi}) is well defined  thanks to this condition either under small norm assumptions on $u'(x)$, or under assumption of decaying of this perturbation faster than some exponent linear in $x$. In addition, it is natural to expect that the proof of unique solvability of (\ref{tphi}) could be performed in analogy to the case of a decaying potential, see~\cite{Wickerhauser,FS,Grinevich0}.

Here, we assume this unique solvability and study the properties of the total Green's function, Jost solutions and scattering data of the perturbed potential $u(x)$, exploiting the fact that their properties are inherited, via equations of the kind (\ref{tphi}), from properties of these objects in the pure $N$ soliton case. This approach was already developed in~\cite{BPPP2002} for the perturbation of the one soliton potential (see also~\cite{VA}, where some class of potentials non decaying in one space direction was considered), while the generic case of the $N$-soliton potential was left open just because the Green's function obeying boundedness condition was unknown. This problem was solved in our article~\cite{ResolventNheat} (preliminary version was given in~\cite{KPIIGreen}), where by means of the extended resolvent approach we derived the Green's function $\Go _{N}(x,y,\bk)$ and presented its properties in detail. In particular, we showed that this function is discontinuous at some points on the $\bk$-plane, determined by the soliton parameters, and we derived a set of auxiliary Green's functions that enable the control of the discontinuities of $\Go _{N}(x,y,\bk)$ and Jost solutions, see Sec.~\ref{2gr} below.

The article is organized as follows. In Secs.\ 2 and 3 we give necessary details for the pure $N$ soliton potential, Jost solutions and Green's function. The Direct problem for the perturbed potential (\ref{tu}) is given in Sec.\ 4. In Sec.\ 5 we investigate properties of the corresponding Jost solution and scattering data and derive the Inverse problem. Time evolution of the scattering data is given in Sec.\ 6. Concluding remarks and some future developments are presented in Sec.\ 7.

\section{Heat operator with multisoliton potential and its Jost solutions}\label{NpJ}

Soliton potentials (see\cite{BPPP2009} and\cite{asympJostKPII,BPPPr2001a,asympKPII,BK,BC2,ChK2,K,equivKPII} for details) are
labeled by two numbers (topological charges) $N_{a}$ and $N_{b}$, which obey condition
\begin{equation}
N_{a},N_{b}\geq 1. \label{nanb}
\end{equation}
Let
\begin{equation}
\No=N_{a}+N_{b},\text{ so that } \No\geq 2. \label{Nnanb}
\end{equation}
We introduce the $\No$ real parameters
\begin{equation}
\kappa_{1}<\kappa_{2}<\ldots <\kappa_{\No},
\label{kappas}
\end{equation}
and functions
\begin{equation}
K_{n}^{}(x)=\kappa_{n}^{}x_{1}^{}+\kappa_{n}^{2}x_{2}^{},\quad n=1,\ldots,\No. \label{Kn}
\end{equation}
Let
\begin{equation}
e^{K(x)}=\diag\{e^{K_{n}(x)}\}_{n=1}^{\No} \label{eK}
\end{equation}
be a diagonal $\No\times{\No}$ matrix, let $\Do$ be a $\No\times {N_{b}}$ real constant matrix, with at least two nonzero maximal minors, and let $\Do^{\,\prime}$ be a constant $N_{a}\times\No$ matrix that, like the matrix $\Do$, has at least two nonzero maximal minors and that is orthogonal to the matrix $\Do$ in the sense that
\begin{equation}
\Do^{\,\prime}\Do=0, \label{dd}
\end{equation}
where the zero in the r.h.s.\ is a $N_{a}\times {N_{b}}$ matrix. Let us also introduce the two ``incomplete
Vandermonde matrices''  ($N_{b}\times\No$ and $\No\times {N_{a}}$, correspondingly)
\begin{equation}
\Vo=\left(\begin{array}{lll}
1 &\ldots & 1\\
\kappa_{1} &\ldots &\kappa_{\No}\\
\vdots &  &\vdots\\
\kappa_{1}^{N_{b}-1} &\ldots &\kappa_{\No}^{N_{b}-1}
\end{array}\right),\qquad
\Vo^{\,\prime}=\left(\begin{array}{lll}
1 &\ldots &\kappa_{1}^{N_{a}-1}\\
\vdots &  &\vdots\\
1 &\ldots &\kappa_{\No}^{N_{a}-1}
\end{array}\right).  \label{W}
\end{equation}
In analogy to (\ref{dd}) these matrices obey the orthogonality condition
\begin{equation}
\Vo\gamma\Vo^{\,\prime}=0,\label{vv}
\end{equation}
where $\gamma $ is the $\No\times\No$ constant, diagonal, real matrix
\begin{equation}
\gamma =\diag\{\gamma_{n}\}_{n=1}^{\No},\qquad\gamma_{n}=
\prod_{\substack{n'=1 \\ n'\neq {n}}}^{\No}(\kappa_{n}-\kappa_{n'})^{-1}. \label{gamma}
\end{equation}

Then, the soliton potential is given by any of two representations
\begin{equation}
u_{N}(x)=-2\partial_{x_{1}}^{2}\log\tau^{}_{N}(x)\equiv-2\partial_{x_{1}}^{2}\log\tau'_{N}(x), \label{ux}
\end{equation}
where the $\tau$-functions are expressed as
\begin{equation}
\tau^{}_{N}(x)=\det\bigl(\Vo e^{K(x)}\Do\bigr),\qquad\tau'_{N}(x)=
\det\left(\Do^{\,\prime}e^{-K(x)}\gamma\Vo^{\,\prime}\right) . \label{tau}
\end{equation}
For the Jost and dual Jost solutions (solutions, respectively, of the heat operator (\ref{heatop}) and its dual) we have
\begin{equation}
\Phi_{N}(x,\bk)=e^{-i\bk x_{1}-\bk^{2}x_{2}}\dfrac{\tau_{\Phi_{N}}^{}(x,\bk)}{\tau_{N}(x)},\qquad
\Psi_{N}(x,\bk)=e^{i\bk x_{1}+\bk^{2}x_{2}}\dfrac{\tau_{\Psi_{N}}^{}(x,\bk)}{\tau_{N}(x)},\label{PhiPsi}
\end{equation}
where $\tau_{\Phi_{N}}^{}(x,\bk)$ and $\tau_{\Psi_{N}}^{}(x,\bk)$ are given in terms of the $\tau$-functions by means of the Miwa shift:
\begin{equation}
\tau_{\Phi_{N}}(x,\bk)=\det\bigl(\Vo e^{K(x)}(\kappa+i\bk)\Do\bigr),\qquad
\tau_{\Psi_{N}}(x,\bk)=\det\left(\Vo\dfrac{e^{K(x)}}{\kappa+i\bk}\Do\right), \label{tauk:1}
\end{equation}
and where we introduced one more $\No\times\No$ diagonal matrix
\begin{equation}
\kappa+i\bk=\diag\{\kappa_{n}+i\bk\}_{n=1}^{\No}. \label{kappadiag}
\end{equation}
Analogous relations are valid in terms of  $\tau'_{\No}$. Thanks to these definitions we have the following conjugation property of the Jost solutions:
\begin{equation}
\ol{\Phi_{N}^{}(x,\bk)}=\Phi_{N}^{}(x,-\ol\bk),\qquad \ol{\Psi_{N}^{}(x,\bk)}=\Psi_{N}^{}(x,-\ol\bk). \label{realPhiPsi}
\end{equation}

In order to study the properties of the potential and the Jost solutions, it is convenient to use the representations for the $\tau$-functions, which follow from Binet--Cauchy formula for the determinant of a product of matrices, i.g.\ for the $\tau$-function we have
\begin{align}
\tau_{N}(x)&=\dfrac{1}{N_{b}!}\sum_{\{n_{i}\}=1}^{\No}\Do(\{n_{i}\})\Vo(\{n_{i}\})
\prod_{l=1}^{N_{b}}e^{K_{n_{l}}(x)}, \label{tauf1}\\
\tau_{\Phi_{N}}(x,\bk)&=\dfrac{1}{N_{b}!}\sum_{\{n_{i}\}=1}^{\No}\Do(\{n_{i}\})\Vo(\{n_{i}\})
\prod_{l=1}^{N_{b}}(\kappa_{n_{l}}+i\bk)e_{}^{K_{n_{l}}(x)},\label{tauf2}\\
\tau_{\Psi_{N}}(x,\bk)&=\dfrac{1}{N_{b}!}\sum_{\{n_{i}\}=1}^{\No}\Do(\{n_{i}\})\Vo(\{n_{i}\})
\prod_{l=1}^{N_{b}}\dfrac{e_{}^{K_{n_{l}}(x)}}{\kappa_{n_{l}}+i\bk}, \label{tauf3}
\end{align}
where we use notation
\begin{equation}
\Vo(\{n_{i}\})=\left|\begin{array}{lll}
1 &\ldots & 1\\
\kappa_{n_{1}} &\ldots &\kappa_{n_{N_{b}}}\\
\vdots &  &\vdots\\
\kappa_{n_{1}}^{N_{b}-1} &\ldots &\kappa_{n_{N_{b}}}^{N_{b}-1}\end{array}\right|,\qquad
\Do(\{n_{i}\})=\left|\begin{array}{ccc}
\Do_{n_{1},1} &\dots &\Do_{n_{1},N_{b}}\\
\vdots &  &\vdots\\
\Do_{n_{N_{b}},1} &\dots &\Do_{n_{N_{b}},N_{b}}\end{array}\right| \label{VDo1}
\end{equation}
for the maximal minors of matrices $\Vo$ and $\Do$ and where $\{n_{i}\}=\{n_{1},\ldots,n_{N_{b}}\}$
stands for not necessary ordered sets of $N_{b}$ indexes from the interval $1,\ldots,\No$.

We recall that the maximal minors of a matrix satisfy the Pl\"{u}cker relation, i.e., for any subsets $\{m_{i}\}$ and $\{n_{i}\}$ of indexes running from $1$ to $\No$ and arbitrary $j\in\{1,\ldots,N_{b}\}$
\begin{align}
&\Do(\{m_{i}\})\Do(\{n_{i}\})=\notag\\
&=\sum_{s=1}^{N_{b}}\Do(m_{1},\ldots,m_{s-1},n_{j},m_{s+1},\ldots,m_{N_{b}})
\Do(n_{1},\ldots,n_{j-1},m_{s},n_{j+1},\ldots,n_{N_{b}}).\label{plu}
\end{align}
The matrices $\Do$ and $\Do'$ are not in one-to-one correspondence with the potential $u(x)$. Indeed, from (\ref{ux}) and (\ref{tau}), one gets that the potential is invariant under the substitutions
\begin{equation}
\Do\rightarrow\Do v,\qquad\Do^{\,\prime}\rightarrow {v'}\Do^{\,\prime}, \label{ddv}
\end{equation}
where $v$ and $v'$ are respectively arbitrary real constant, nonsingular $N_{b}\times {N_{b}}$ and $N_{a}\times{N_{a}}$ matrices. Thus under condition (\ref{kappas}), the ($N_{a},N_{b}$)-soliton potential is parameterized by a point on the Grassmanian $\text{Gr}_{N_{b},\No}$ if the first representation in (\ref{tau}) is used, or by a point on its dual parametrization (i.e., the Grassmanian $\text{Gr}_{N_{a},\No}$) if the second representation in (\ref{tau}) is used.

Regularity of the potential $u(x)$ on the $x$-plane is equivalent to the absence of zeros of $\tau (x)$. It is clear that it is enough to impose the condition that the matrix $\Do$ is Totally Non Negative (TNN), i.e., that
\begin{equation}
\Do(n_{1},\ldots,n_{N_{b}})\geq 0,\quad\text{for all}\quad 1\leq n_{1}<\ldots <n_{N_{b}}\leq\No. \label{D}
\end{equation}
It is known that in the classification of the TNN matrices a special role is played by the solid minors \cite{FZ}, i.e., minors of the kind $\Do(n,\ldots,n+N_{b}-1)$ for all $n=1,\ldots,\No$ where we assume that the indexes are defined mod$\,\No$, so that thanks to (\ref{Nnanb}), say, $n+N_{b}=n-N_{a}$ for $n>N_{a}$. In~\cite{asympKPII} we proved that if the coefficients
\begin{equation}
z_{n}=V(n,\ldots,n+N_{b}-1)\Do(n,\ldots,n+N_{b}-1), \label{zn}
\end{equation}
are strictly positive for all $n=1,\ldots,\No$ (again with indexes defined mod$\,\No$) the function $\tau (x)$ has the following asymptotic behavior
\begin{align}
& x\overset{r_{n}}{\longrightarrow}\infty : & &\tau (x)=
\bigl(z_{n}+z_{n+1}e_{}^{K_{N_{b}+n}(x)-K_{n}(x)}+o(1)\bigr)
\exp\Biggl(\sum_{l=n}^{n+N_{b}-1}K_{l}(x)\Biggr), \label{tau:rn}\\
& x\overset{\sigma_{n}}{\longrightarrow}\infty : & &\tau (x)=\bigl(z_{n}+o(1)\bigr)\exp\Biggl(\sum_{l=n}^{n+N_{b}-1}K_{l}(x)\Biggr).
\label{tau:sn}
\end{align}
Here $r_n$ denotes the asymptotic rays on the $x$-plane
\begin{equation}
r_{n}:\qquad\left\{\begin{array}{l}
x_{1}+(\kappa_{n}+\kappa_{n+N_{b}})x_{2}\quad\text{bounded}\\
(\kappa_{n+N_{b}}-\kappa_{n})x_{2}\rightarrow -\infty .
\end{array}\right.,\quad n=1,\ldots,\No, \label{rn}
\end{equation}
so that there are $N_{a}$ rays in the direction $x_{2}\to-\infty $ and $N_{b}$ rays in the direction $x_{2}\to+\infty$. As well, $\sigma_{n}$ denotes the sector swept out by rotating anticlockwise the ray $r_{n}$ up to the ray $r_{n+1}$. These
sectors are nonintersecting and cover the whole $x$-plane with the exception of rays.

It is clear that the condition
\begin{equation}
z_{n}>0 \label{zn0}
\end{equation}
is sufficient for having nonsingular asymptotics of the potential. In the case of a TNN matrix, this condition is equivalent\cite{Private} to the condition that all maximal minors of the matrix $\Do$ are positive, i.e., $\Do$ is a totally positive (TP) matrix.

A special role in the construction below is played by the values $\Phi_{N}(x,i\kappa_{n})$ of the $\Phi_{N}(x,\bk)$ at points $i\kappa_n$ and  the residuals of  $\Psi_{N}(x,\bk)$ at the same points. In terms of these discrete values we define a $\No$-row
\begin{equation}
\Phi_{N,\kappa}(x)=\{\Phi_{N,1}(x),\ldots,\Phi_{N,\No}(x)\}, \quad\Phi_{N,n}(x)=\Phi_{N}(x,i\kappa_{n}^{}),\label{Phik}
\end{equation}
and a $\No$-column
\begin{equation}
\Psi_{N,\kappa}(x)=\{\Psi_{N,1}(x),\ldots,\Psi_{N,\No}(x)\}^{\text{T}},\quad \Psi_{N,n}(x)=-i\res_{\bk=i\kappa_{n}}\Psi(x,\bk),  \label{Psik}
\end{equation}
$n=1,\ldots,\No$. In\cite{asympJostKPII} (in a bit different notation) we proved that
\begin{align}
&\Phi_{N,\kappa}(x)\Do=0,\label{Phid}\\
&\Do^{\,\prime}\Psi_{N,\kappa}(x)=0,\label{dPsi}
\end{align}
and that the Jost solutions obey the Hirota bilinear identity
\begin{equation}
\sum_{n=1}^{\No}\Phi_{N,n}(x)\Psi_{N,n}(x')=0.\label{sumPhiPsi}
\end{equation}
Thanks to (\ref{realPhiPsi}) we have that
\begin{equation}
\ol{\Phi_{N,n}(x)}=\Phi_{N,n}(x),\qquad \ol{\Psi_{N,n}(x')}=\Psi_{N,n}(x'),\label{realPhiPsin}
\end{equation}
for all $n$.

In~\cite{BPPPr2001a,asympJostKPII,equivKPII,asympKPII} detailed properties of the Jost solutions and their discrete values are given. Here we mention only that the functions $e^{i\bk{x}_1+\bk^{2}x_2}\Phi_{N}(x,\bk)$ and
$e^{-i\bk{x}_1-\bk^{2}x_2}\Psi_{N}(x,\bk)$ have bounded asymptotics on the $x$-planes, function $e^{i\bk{x}_1+\bk^{2}x_2}\Phi_{N}(x,\bk)$ is a polynomial with respect to $\bk$ of order $N_b$, and function
$e^{-i\bk{x}_1-\bk^{2}x_2}\Psi_{N}(x,\bk)$ is a meromorphic function of $\bk$ with simple poles at all $\bk=i\kappa_n$ and of order $-N_b$ at infinity. More exactly, we have the following normalization
\begin{equation}
\lim_{\bk\rightarrow\infty}(i\bk)^{-N_{b}}e^{i\bk x_{1}+\bk^{2}x_{2}}\Phi_{N}(x,\bk)=1,\qquad
\lim_{\bk\rightarrow\infty}(i\bk)^{N_{b}}e^{-i\bk x_{1}-\bk^{2}x_{2}}\Psi_{N}(x,\bk)=1, \label{asymptk2}
\end{equation}
so that the potential can be reconstructed as
\begin{align}
u_{N}^{}(x)& =-2\lim_{\bk\rightarrow\infty}(i\bk)^{-N_{b}+1}\partial_{x_{1}}
\bigl(e^{i\bk x_{1}+\bk^{2}x_{2}}\Phi_{N}(x,\bk)\bigr)\equiv \notag\\
&\equiv 2\lim_{\bk\to\infty}(i\bk)^{N_{b}+1}\partial_{x_{1}}\bigl(e^{-i\bk x_{1}-\bk^{2}x_{2}}\Psi_{N}(x,\bk)\bigr). \label{asymptk3}
\end{align}

Analyticity properties of the Jost solutions with respect to $\bk$ impose conditions on their discrete values. Indeed, taking notations (\ref{Kn}), (\ref{eK}), and (\ref{gamma}) into account, we can write that
\begin{equation}
e^{i\bk x_{1}+\bk^{2}x_{2}}_{}\Phi_{N}(x,\bk)=(-1)^{\No-1}\prod\limits_{m=1}^{\No}(\kappa_{m}+i\bk)\sum_{n=1}^{\No}
\dfrac{\gamma_{n}e^{-K_{n}(x)}\Phi_{N,n}(x)}{\kappa_{n}+i\bk}.\label{ipN}
\end{equation}
On the other side we know that the l.h.s.\ is a polynomial of the order $N_{b}$, normalized according to (\ref{asymptk2}). Thus, since all higher powers in the above equality must cancel out, we obtain that
\begin{equation}
\Phi_{N,\kappa}(x){\gamma}e^{-K(x)}\Vo^{\,\prime}=\bigl(\underbrace{0,\ldots,0,(-1)^{N_{b}}}_{N_{a}}\bigr). \label{ipN:Phi}
\end{equation}
An analogous formula for $\Psi_{N,\kappa}(x)$ can be derived directly, in a similar way, or by recalling (see~\cite{asympJostKPII}) that
\begin{equation}
\Psi_{N}(x,\bk)=\dfrac{[\Phi_{N}(-x,\bk)]}{\prod\limits_{n=1}^{\No}(\kappa_{n}+i\bk)}, \label{Psi_Phi}
\end{equation}
where square brackets in the r.h.s.\ means that one has to perform in $\Phi_{N}(-x,\bk)$ the substitutions $\Do\rightarrow\gamma\Do^{\prime\text{T}}$, $N_{a}\leftrightarrow{N}_{b}$, while the $\kappa_{n}$'s remain unchanged. Precisely, we have from (\ref{Psi_Phi}) that $\Psi_{n}(x)=(-1)^{\No}\gamma_{n}[\Phi_{N,n} (-x)]$, and, then, thanks to (\ref{ipN:Phi}) we get
\begin{equation}
\Vo{e}^{K(x)}\Psi_{N,\kappa}(x)=\bigl(\underbrace{0,\ldots,(-1)^{N_{b}}}_{N_b}\bigr)^{\text{T}}. \label{ipN:Psi}
\end{equation}
Equalities (\ref{Phid}) and (\ref{ipN:Phi}) define the discrete values $\Phi_{N,n}(x)$ and then the Jost solution $\Phi_{N}(x,\bk)$ itself by (\ref{ipN}). The same is valid, of course, for the dual Jost solution $\Psi_{N}(x,\bk)$, where the defining equations are (\ref{dPsi}) and (\ref{ipN:Psi}). The potential $u_{N}(x)$ can be reconstructed by (\ref{asymptk3}).

In order to simplify relations below, we introduce operatorial notation of the kind
\begin{equation}
\overrightarrow{\Lo}_{N}\Phi_{N}(\bk)=0,\qquad\Psi_{N}(\bk)\overleftarrow{\Lo}_{N}=0, \label{JostandDual}
\end{equation}
for the heat equation and its dual and
\begin{equation}
\overrightarrow{\Lo}_{N}\Go_{N}(\bk)=\Go_{N}(\bk)\overleftarrow{\Lo}_{N}=I,\label{GreenN}
\end{equation}
for the equation (\ref{green}) and its dual. Thus $\overrightarrow{\Lo}_{N}$ denotes the operator $\Lo_{N}$ applied to the $x$ variable of the kernel $\Go_{N}(x,x',\bk)$ and $\overleftarrow{\Lo}_{N}$ denotes the operator dual to $\Lo_{N}$ (i.e., $\Lo^{\text{d}}_{N}$)  applied to the $x'$ variable of the same kernel.

\section{Green's functions of the pure soliton potential}\label{2gr}
\subsection{Green's function $g_{N}(x,x')$}\label{gN}

In \cite{ResolventNheat} by means of the extended resolvent approach we derived the Green's function of the heat operator with generic $N$-soliton potential obeying condition (\ref{zn0}) (for a preliminary version see \cite{KPIIGreen}). In terms of the Jost solutions it can be written in the form
\begin{align}
\Go_{N}(x,x',\bk)& =-\dfrac{\sgn(x_{2}-x_{2}')}{2\pi }\int \!\!ds\,\theta
\bigl((s^{2}-\bk^{2}_{\Re})(x^{}_{2}-x_{2}')\bigr)\Phi_{N}(x,s+i\bk_{\Im})\Psi_{N}(x^{\prime},s+i\bk_{\Im})-\nonumber\\
&-\theta(x'_{2}-x^{}_{2})\sum_{n=1}^{\No}\theta (\bk_{\Im}-\kappa _{n})\Phi_{N,n}(x)\Psi_{N,n}(x').\label{g3}
\end{align}
We proved there that function $G_{N}(x,x'\bk')$ defined in (\ref{Green}) is bounded with respect to the variables $x,x'\in\Rs^{2}$ and $\bk\in\Cs$ and has finite limits at infinity. We also derived the auxiliary Green's function
\begin{align}
g_{N}(x,x')&=-\dfrac{\theta(x_{2}-x_{2}')}{2\pi}\Biggl\{\int d\alpha\,e_{}^{-i\alpha(x^{}_1-x'_1)-\alpha^2(x^{}_2-x'_2)}+\nonumber\\
&+i\sum_{n=1}^{\No}\Phi_{N,n}^{}(x)\Psi^{}_{N,n}(x')\int \dfrac{d\alpha}{\alpha}\,
e_{}^{-i\alpha(x^{}_1-x'_1+2\kappa_{n}(x^{}_2-x'_2))-\alpha^{2}(x^{}_2-x'_2)}\Biggr\},\label{g4}
\end{align}
where integrals are understood in the sense of the principal value (in \cite{ResolventNheat} this function was denoted as $\Go^{+}$). Function $\Go_{N}(x,x',\bk)$ is continuous with respect to the variable $\bk\in\Cs$ with exception of the points $\bk=i\kappa_{1},\ldots,i\kappa_{\No}$. Discontinuities at this points can be described by means of the auxiliary Green's function (see below). This function can be calculated explicitly as
\begin{align}
&g_{N}(x,x') =-\dfrac{\theta (x_{2}^{}-x_{2}')}{2\sqrt{\pi (x_{2}^{}-x_{2}')}}
\exp\biggl(-\dfrac{(x_{1}^{}-x'_{1})^{2}}{4(x_{2}^{}-x_{2}')}\biggr)- \notag\\
&\qquad-\dfrac{\theta (x_{2}^{}-x_{2}')}{2}\sum_{n=1}^{\No}\Phi_{N,n}(x)\Psi_{N,n}(x')
\biggl[\erf\biggl(\dfrac{x_{1}^{}-x_{1}'+2\kappa_{n}(x_{2}^{}-x_{2}')}
{\sqrt{x_{2}^{}-x_{2}'}}\biggr)-1\biggr], \label{1_6}
\end{align}
where the error function is defined by
\begin{equation}
\erf(x)=\dfrac{2}{\sqrt{\pi}}\int\limits_{0}^{x}dy\,e^{-y^{2}}, \label{1_7}
\end{equation}
so that
\begin{equation}
\lim_{x\rightarrow\infty}\erf(x)=1,\qquad\lim_{x\rightarrow\infty}[\erf(x)-1]xe^{x^{2}}=\dfrac{-1}{\sqrt{\pi}}. \label{1_8}
\end{equation}
Taking into account (\ref{sumPhiPsi}), we subtracted $1$ in the square brackets of the r.h.s.\ for making explicit the good behavior at large $x$ of the last term in the r.h.s.\ thanks to the first limit in (\ref{1_8}).  To specify this behavior we consider
\begin{align}
& e_{}^{\bk_{\Im}(x_{1}'-x_{1}^{})+\bk_{\Im}^{2}(x_{2}'-x_{2}^{})}g_{N}^{}(x,x')=\notag\\
&\quad =-\dfrac{\theta (x_{2}^{}-x_{2}')}{2\sqrt{\pi(x_{2}^{}-x_{2}')}}\exp\biggl(-\dfrac{(x_{1}^{}-x'_{1}+2\bk_{\Im}(x_{2}^{}-x_{2}'))^{2}}{4(x_{2}^{}-x_{2}')}\biggr)- \notag\\
&\quad-\dfrac{\theta (x_{2}^{}-x_{2}')}{2}\sum_{n=1}^{\No}e^{K_n(x)}\Phi_{N,n}(x)e^{-K_n(x')}\Psi_{N,n}(x')
e_{}^{(\kappa_{n}-\bk_{\Im})(x_{1}^{}-x_{1}'+2\kappa_{n}(x_{2}^{}-x'_{2}))}\times \notag\\
&\quad\times e_{}^{-(\kappa_{n}-\bk_{\Im})^{2}(x_{2}^{}-x_{2}')}
\biggl[\erf\biggl(\dfrac{x_{1}^{}-x_{1}'+2\kappa_{n}(x_{2}^{}-x_{2}')}{\sqrt{x_{2}^{}-x_{2}'}}\biggr)-1\biggr].\label{1_11}
\end{align}
When $x\to\infty$ the first term in the r.h.s.\  is decaying in all $x$ directions and for any $\bk_{\Im}$ at least as
$(x_{2}-x_{2}')^{-1/2}$. Let us consider the second term taking into account that all $e^{K_n(x)}\Phi_{N,n}(x)$ and
$e^{-K_n(x')}\Psi_{N,n}(x')$ are bounded for all $x$ and have finite limits at infinity.  The first exponent under summation sign can grow linearly with respect to the combination $x_{1}^{}-x_{1}'+2\kappa_{n}(x_{2}^{}-x_{2}')$. But for generic situation this is depressed by the behavior of the second exponent and $\erf$ in the second equality of (\ref{1_8}). Only the special case, in which the asymptotic behavior of $x$ is such that for some $n$ the argument of $\erf$ is bounded, is left. But then the first exponential factor is depressed by the second one if $\bk_{\Im}\neq\kappa_{n}$. If $\bk_{\Im}=\kappa_{n}$, then both these exponents are absent and again we have boundedness of the r.h.s.\ in (\ref{1_11}). This proves that for any $\bk_{\Im}\in\Rs$ the product
\begin{equation}
e_{}^{-\bk_{\Im}(x_{1}^{}-x_{1}')-\bk_{\Im}^{2}(x_{2}^{}-x_{2}')}g_{N}^{}(x,x'), \label{1_9}
\end{equation}
is bounded. Let us mention also that the Green's function $g_{N}$ is real,
\begin{equation}
\ol{g_{N}^{}(x,x')}=g_{N}^{}(x,x'), \label{real_g}
\end{equation}
as follows from (\ref{realPhiPsi}), (\ref{realPhiPsin}), and (\ref{1_7}).

\subsection{Green's function $\Go_{N}(x,x',\bk)$}

By definition function  $\Go_{N}(x,x',\bk)$ obeys
\begin{equation}
\ol{\Go_{N}^{}(x,x',\bk)}=\Go_{N}^{}(x,x',-\ol\bk)=\Go_{N}^{}(x,x',\bk), \label{real_G}
\end{equation}
thanks to (\ref{realPhiPsi}), (\ref{realPhiPsin}), and in terms of distributions we have by  (\ref{g3})
\begin{equation}
\dfrac{\partial\Go_{N}^{}(\bk)}{\partial\ol{\bk}}=\dfrac{\sgn\bk_{\Re}^{}}{2\pi}\,\Phi_{N}(-\ol{\bk})\otimes
\Psi_{N}(-\ol{\bk}),\label{dG}
\end{equation}
where for shortness we omit space variables and use the notation of the direct product in the standard way: $\bigl(\Phi_{N}(-\ol{\bk})\otimes
\Psi_{N}(-\ol{\bk})\bigr)(x,x')\equiv\Phi_{N}(x,-\ol{\bk})\Psi_{N}(x',-\ol{\bk})$. In \cite{ResolventNheat} we proved that the Green's function (\ref{g3}) can be written in the form
\begin{align}
&\Go_{N}(x,x',\bk)=g_{N}(x,x')+ \notag\\
&\qquad+\dfrac{1}{2\pi}\int_{-\left\vert\bk_{\Re}\right\vert}^{\left\vert\bk_{\Re}\right\vert}d\alpha\,
\left[\Phi_{N}(x,\alpha+i\bk_{\Im})\Psi_{N}(x',\alpha+i\bk_{\Im})-\sum_{m=1}^{\No}\dfrac{\Phi_{N,m}(x)
\Psi_{N,m}(x')}{\bk_{\Im}-\kappa_{m}-i\alpha}\right]+\notag\\
&\qquad+\sum_{m=1}^{\No}a(\bk-i\kappa_{m})\Phi_{N,m}(x)\Psi_{N,m}(x'), \label{Gg}
\end{align}
where we introduced the function
\begin{equation}
a(\bk)=\dfrac{1}{\pi}\text{arccot}\dfrac{\bk_{\Im}}{|\bk_{\Re}|}.\label{ak}
\end{equation}
Function $a(\bk)$ satisfies the following relations
\begin{align}
&\ol{a(\bk)}=a(\bk)=a(-\ol{\bk}),\qquad 0\leq {a(\bk)}\leq 1,\label{ak1}\\
& a(ie^{i\alpha}|\bk|)=\dfrac{|\alpha |}{\pi},\quad -\pi <\alpha <\pi,\label{ak2}
\end{align}
Since the integrand in (\ref{Gg}) has no singularities, we see that, indeed, the Green's function $\Go_{N}(x,x',\bk)$ is discontinuous at all points $\bk=i\kappa_{n}$ and has finite limits at these points depending on the limiting procedure (see (\ref{ak2})). In order to control these discontinuities it is convenient to introduce an auxiliary Green's function that in the case of the heat equation (see\cite{ResolventNheat}) coincides with the value of $\Go_{N}(x,x',\bk)$ at $\bk_{\Im}=0$. Since the term in the second line of (\ref{Gg}) tends to zero when $\bk_{\Re}\to0$ we have by (\ref{Gg}):
\begin{equation}
\Go_{N}(x,x',i\bk_{\Im})=g_{N}(x,x')+\sum_{m=1}^{\No}\Phi_{N,m}(x)\Psi_{N,m}(x')
\theta (\kappa_{m}-\bk_{\Im}), \label{Gg1}
\end{equation}
which is defined for any value of $\bk_{\Im}\neq\kappa_{n}$, $n=1,\ldots,\No$. It is discontinuous at all points $\bk_{\Im}=\kappa_{1},\ldots,\kappa_{\No}$ and piecewise constant function of $\bk_{\Im}$ otherwise. In the case where $\bk_{\Im}<\kappa_{1}$, or $\bk_{\Im}>\kappa_{\No}$ we have thanks to (\ref{sumPhiPsi}) that $\Go_{N}(x,x',i\bk_{\Im})=g_{N}(x,x')$ and
\begin{equation}
\lim_{\bk_{\Im}\to\kappa_{n}-0}\Go_{N}(i\bk_{\Im})=\lim_{\bk_{\Im}\to\kappa_{n}+0}\Go_{N}(i\bk_{\Im})+
\Phi_{N,n}\otimes\Psi _{N,n}.  \label{G+-}
\end{equation}
Because of (\ref{realPhiPsin}) and (\ref{real_g}) this function is real
\begin{equation}
\ol{\Go^{}_{N}(i\bk_{\Im})}=\Go^{}_{N}(i\bk_{\Im}),\label{realGI}
\end{equation}
and taking its discontinuity into account we denote
\begin{equation}
\Go_{N,n}^{}=\lim_{\bk_{\Im}\to\kappa_{n}+0}\lim_{\bk_{\Re}\to0}\Go_{N}(\bk),\quad n=1,\ldots,\No.\label{GNn}
\end{equation}
Then by (\ref{Gg1}) we have that
\begin{equation}
\Go_{N,n}^{}=g_{N}^{}+\sum_{m=n+1}^{\No}\Phi_{N,m}(x)\Psi_{N,m}(x')
\equiv{g}_{N}^{}-\sum_{m=1}^{n}\Phi_{N,m}(x)\Psi_{N,m}(x'), \label{Gg2}
\end{equation}
where indexes are defined $\mod\No$, as it was mentioned after (\ref{D}), and where the second equality follows by (\ref{sumPhiPsi}). Thus
\begin{equation}
\Go_{N,n-1}=\Go_{N,n}+\Phi_{N,n}\otimes\Psi _{N,n}, \label{GNn-1}
\end{equation}
and then for any $\bk$ in a neighborhood of a point $i\kappa_{n}$ and any $n=1,\ldots,\No$ we have:
\begin{equation}
\Go_{N}^{}(\bk)=\Go_{N,n}^{}+a(\bk-i\kappa_{n})\Phi_{N,n}\otimes\Psi_{N,n}+o(1),
\quad\bk\cong i\kappa_{n}.\label{GNa}
\end{equation}

\section{Perturbation of $N$-soliton potential}
\subsection{Main definitions}

Let us now consider the heat operator (\ref{heatop}) with an $N$-soliton potential perturbed by adding to it a real function $u'(x)$, smooth and rapidly decaying at space infinity, see (\ref{tu}), i.e.,
\begin{equation}
\Lo(x,\partial_{x})=\Lo_{N}^{}(x,\partial_{x})-u'(x).\label{LLN}
\end{equation}
Below we use the operatorial notation $\Lo=\Lo_{N}^{}-U'$, where $U'$ denotes the multiplication operator with kernel
\begin{equation}
U'(x,x')=u'(x)\delta (x-x').\label{Uu}
\end{equation}
Thus, we write the integral equation (\ref{tphi}) for the Jost solution of the heat operator $\Lo$ in (\ref{LLN}), and for its dual, as follows
\begin{equation}
\Phi (\bk)=\Phi_{N}^{}(\bk)+\,\Go_{N}^{}(\bk)U'\Phi (\bk),\quad\Psi (\bk)=\Psi_{N}^{}(\bk)+\Psi (\bk)\,U'\Go_{N}^{}(\bk).\label{dpJ}
\end{equation}
In fact we introduce and study here a more general object: the total Green's function $\Go(x,x',\bk)$ defined for any value of the complex spectral parameter $\bk$ by means of one of the following integral equations:
\begin{equation}
\Go(\bk)=\Go_{N}^{}(\bk)+\,\Go_{N}^{}(\bk)U'\Go(\bk),\qquad\Go(\bk)=\Go_{N}^{}(\bk)+\,\Go(\bk)U'\Go_{N}^{}(\bk). \label{dpG}
\end{equation}

As it was mentioned in the Introduction, we assume here that these equations are uniquely solvable and give the same solution at least for smooth $u'(x)$ obeying some small norm conditions, or decaying fast enough at space infinity, e.g., faster than some decaying exponent linear in $x$. We also assume that this function inherits properties of the pure solitonic Green's function $\Go_{N}^{}(\bk)$ by means of (\ref{dpG}). In particular, $\Go(x,x',\bk)$ obeys the differential equations
\begin{equation}
\overrightarrow{\Lo}\Go(\bk)=\Go(\bk)\overleftarrow{\Lo}=I\label{deG}
\end{equation}
and, thanks to (\ref{GreenN}), boundedness property (\ref{Green}). This function is continuous, continuously differentiable with respect to the complex spectral parameter $\bk$, for all values of this parameter different from $i\kappa_{1},\ldots,i\kappa_{\No}$. Because of (\ref{real_G}) and reality of the potential $u'$ we have the conjugation property
\begin{equation}
\ol{\Go(x,x',\bk)}=\Go(x,x',-\ol{\bk})=\Go(x,x',\bk). \label{realG}
\end{equation}
In terms of the Green's function we define the Jost and dual Jost solutions as
\begin{equation}
\Phi (\bk)=\Go(\bk)\overleftarrow{\Lo}_{N}^{}\Phi_{N}^{}(\bk),\quad
\Psi (\bk)=\Psi_{N}^{}(\bk)\overrightarrow{\Lo}_{N}^{}\Go(\bk),\label{2-6}
\end{equation}
where, say, the first equation is a short form for
\begin{equation*}
\Phi(x,\bk)=\int{dx'}\{\Lo_{N}^{\text{d}}(x',\partial_{x'})\Go(x,x',\bk)\}\Phi_{N}(x',\bk).
\end{equation*}
Then, thanks to (\ref{LLN}), (\ref{JostandDual}), and (\ref{GreenN}), the differential equations
\begin{equation}
\overrightarrow{\Lo}\Phi (\bk)=0,\qquad\Psi (\bk)\overleftarrow{\Lo}=0, \label{deJ}
\end{equation}
hold and, therefore, we indeed can consider $\Phi(x,\bk)$ and $\Psi(x,\bk)$ as the generalization of the Jost solutions to the case where
the perturbation $u'(x)$ is different from zero. Because of (\ref{LLN}) we can write (\ref{2-6}) also in the form
\begin{align}
\Phi (x,\bk)&=\Phi_{N}^{}(x,\bk)+\int dx'\,\Go(x,x',\bk)u'(x')\Phi_{N}^{}(x',\bk),\label{explPhi}\\
\Psi (x,\bk)&=\Psi_{N}^{}(x,\bk)+\int dx'\,\Go(x',x,\bk)u'(x')\Psi_{N}^{}(x',\bk),\label{explPsi}
\end{align}
Thanks to the above assumptions both these solutions are continuous, continuously differentiable function of  $\bk\in\Cs$ for all values of this parameter different from $i\kappa_{1},\ldots,i\kappa_{\No}$. Because of the asymptotic conditions (\ref{Green}) and (\ref{asymptk2}) we have that
\begin{equation}
\lim_{\bk\rightarrow\infty}(i\bk)^{-N_{b}}e^{i\bk x_{1}+\bk^{2}x_{2}}\Phi(x,\bk)=1,\qquad
\lim_{\bk\rightarrow\infty}(i\bk)^{N_{b}}e^{-i\bk x_{1}-\bk^{2}x_{2}}\Psi(x,\bk)=1, \label{Asymptk}
\end{equation}
and thanks to the reality of the perturbation $u'(x)$ and (\ref{realG}) we have the conjugation properties
\begin{equation}
\ol{\Phi(x,\bk)}=\Phi(x,-\ol\bk),\qquad \ol{\Psi(x,\bk)}=\Psi(x,-\ol\bk). \label{RealPhiPsi}
\end{equation}

\subsection{Behavior of the Green's function at points of discontinuity}

The Green's function $\Go_{N}(x,x',\bk)$ has a well defined limit at $\bk_{\Re}=0$. Correspondingly, the limit $\Go(i\bk_{\Im })$ at $\bk_{\Re}=0$ of the total Green's function obeys the integral equations $\Go(i\bk_{\Im })=\Go_{N}(i\bk_{\Im })+\Go_{N}(i\bk_{\Im })U'\Go(i\bk_{\Im})$ and
$\Go(i\bk_{\Im })=\Go_{N}(i\bk_{\Im })+\Go(i\bk_{\Im })U'\Go_{N}(i\bk_{\Im})$, obeys the differential equations $\overrightarrow{\Lo}\Go(i\bk_{\Im})=\Go(i\bk_{\Im})\overleftarrow{\Lo}=I$, and is a real, piecewise constant function of $\bk_{\Im}$.
Like $\Go_{N}(i\bk_{\Im})$ it has finite limits from the right and from the left at points $\bk_{\Im}=\kappa_{n}$. In analogy to (\ref{GNn}) we denote
\begin{equation}
\Go_{n}^{}=\lim_{\bk_{\Im}\to\kappa_{n}+0}\Go(i\bk_{\Im}),\quad n=1,\ldots,\No,\label{GGk}
\end{equation}
that by (\ref{dpJ}) obeys the integral equations
\begin{equation}
\Go_{n}^{}=\Go^{}_{N,n}+\Go^{}_{N,n}U'\Go_{n}^{},\qquad
\Go_{n}^{}=\Go^{}_{N,n}+\Go_{n}^{}U'\Go^{}_{N,n}. \label{dpGn}
\end{equation}
These values of the Green's function can be related by means of the ``dressed'' version of equation~(\ref{GNn-1}). Say, by the first equation above we get
\begin{equation*}
\Go_{n-1}^{}-\Go_{n}^{}=[\Go^{}_{N,n-1}-\Go^{}_{N,n}](I+U'\Go_{n}^{})+\Go^{}_{N,n-1}U'[\Go_{n-1}^{}-\Go^{}_{n}].
\end{equation*}
Thanks to (\ref{LLN}) and (\ref{dpGn}) we can insert here $I+U'\Go_{n}=\overrightarrow{\Lo}_{N}\Go_{n}$. Then using (\ref{GNn-1}) for the difference $\Go^{}_{N,n-1}-\Go^{}_{N,n}$ we derive
\begin{equation*}
\Go_{n-1}^{}-\Go_{n}^{}=
\Phi_{N,n}\otimes\Psi_{N,n}\overrightarrow{\Lo}_{N}^{}\Go_{n}^{}+\Go^{}_{N,n-1}U'[\Go_{n-1}^{}-\Go^{}_{n}].
\end{equation*}
We see that the difference $\Go_{n-1}-\Go_{n}$ obeys an integral equation with the same kernel as in (\ref{dpGn}) (for $n\to{n}-1$) but with a different inhomogeneous term. Thanks to the equality $\Go_{N,n-1}\overleftarrow{\Lo}_{N}=I$ this term can be obtained from $\Go^{}_{N,n-1}$ by applying operation $\overleftarrow{\Lo}_{N}\Phi_{N,n}\otimes\Psi_{n}^{}$ to (\ref{dpGn}) from the right. Thus under condition of unique solvability of this equation we arrive to
\begin{equation}
\Go_{n-1}^{}-\Go_{n}^{} =\Go^{}_{n-1}\overleftarrow{\Lo}_{N}\Phi_{N,n}\otimes\Psi_{n}^{},\label{diffG}
\end{equation}
where we introduced the auxiliary solutions of the heat operator and its dual (cf.\  (\ref{GNn})):
\begin{align}
\Phi_{n}^{}&=\Go_{n}^{}\overleftarrow{\Lo}_{N}\Phi_{N,n}, \label{Phin1}\\
\Psi_{n}^{}& =\Psi_{N,n}\overrightarrow{\Lo}_{N}\Go_{n}^{},\label{Psin1}
\end{align}
$n\in[1,\No]$, which are real and satisfy $\overrightarrow{\Lo}\Phi_{n}=0=\Psi_{n}\overleftarrow{\Lo}$. They can be considered as the discrete values of the Jost solutions, if we notice that thanks to (\ref{2-6}) and (\ref{GGk}) we have the following generalizations of (\ref{Phik}) and (\ref{Psik}):
\begin{align}
\Phi_{n}^{}(x)&=\lim_{\bk_{\Im}\to\kappa_{n}+0}\Phi(x,i\bk_{\Im}), \label{Phin2}\\
\Psi_{n}^{}(x)& =\lim_{\bk_{\Im}\to\kappa_{n}+0}(\bk_{\Im}-\kappa_{n})\Psi(x,i\bk_{\Im}).\label{Psin2}
\end{align}

The first factor in the r.h.s.\ of (\ref{diffG}) is not of the kind (\ref{Phin1}), so we apply operation $\overleftarrow{\Lo}_{N}\Phi_{N,n}$ to (\ref{diffG}) from the right, that gives
\begin{equation}
\bigl(\Go^{}_{n-1}\overleftarrow{\Lo}_{N}\Phi_{N,n}\bigr)(x)=\dfrac{\Phi_{n}^{}(x)}{1-c_{n}},\label{c1} 
\end{equation}
where we introduced the constants
\begin{equation}
c_{n}^{}=\bigl(\Psi_{n}^{}\overleftarrow{\Lo}_{N}\Phi_{N,n}^{}\bigr),\quad n=1,\ldots,\No.\label{cn}
\end{equation}
Thanks to (\ref{LLN}), (\ref{deJ}), and (\ref{Psin1}) it can be written in the following different forms
\begin{equation}
c_{n}^{}=\bigl(\Psi_{N,n}^{}\overrightarrow{\Lo}_{N}\Go_{n}^{}\overleftarrow{\Lo}_{N}\Phi_{N,n}^{}\bigr)\equiv
\bigl(\Psi_{N,n}^{}\overrightarrow{\Lo}_{N}\Phi_{n}^{}\bigr)\equiv
\bigl(\Psi_{N,n}^{}U'\Phi_{n}^{}\bigr)\equiv\bigl(\Psi_{n}^{}U'\Phi_{N,n}^{}\bigr),\label{cn2} 
\end{equation}
where, say, the last equality is a short form for
\begin{equation}
c_{n}^{}=\int dx\,u'(x)\Psi_{n}^{}(x)\Phi_{N,n}^{}(x).\label{cn3}
\end{equation}
Because of the boundedness property (\ref{Asymptk}) and thanks to (\ref{Kn}) and (\ref{Phin2}) we see that in analogy with the unperturbed $N$-soliton case products
\begin{equation}
e^{-K_{n}(x)}\Phi_{n}^{}(x)\quad\text{and }\quad e^{K_{n}(x)}\Psi_{n}^{}(x)\quad
\text{are bounded for all }x\label{Phinx}
\end{equation}
and have finite limits at $x$-infinity. Thus, the integral in (\ref{cn3}) exists under some small norm condition on $u'(x)$, as discussed above. Then since both sides of (\ref{c1}) exist and $\Phi_{n}(x)$ cannot be identical zero, we get that $c_{n}\neq1$, and thus
\begin{equation}
c_{n}<1, \label{c4}
\end{equation}
because in the limit of the zero perturbation, $u'(x)\to0$, all these constants tend to zero. It is also clear that all these constants are real.

Finally, by (\ref{diffG}) and (\ref{c1}) we get the following generalization of (\ref{GNn-1}) for the perturbed case:
\begin{equation}
\Go_{n-1}=\Go_{n}+\dfrac{\Phi_{n}\otimes\Psi _{n}}{1-c_n}, \label{Gn-1}
\end{equation}
so that, summing over $n$ (under $\mod\No$-condition), we get
\begin{equation}
\sum_{n=1}^{\No}\dfrac{\Phi_{n}\otimes\Psi _{n}}{1-c_n}=0, \label{SumPhiPsi}
\end{equation}
i.e., a generalization of (\ref{sumPhiPsi}).

On the other side, summation of (\ref{Gn-1}) for some interval of indexes in $[1,\No]$ gives the equality
\begin{equation}
\Go_{m}^{}=\Go_{n}+\sum_{l=m+1}^{n+\No}\dfrac{\Phi_{l}\otimes\Psi_{l}}{1-c_l},
\label{GmGn}
\end{equation}
that under $\mod\No$-condition is valid for any $m,n\in[1,\No]$.

In analogy to the above we can consider the difference $\Go(\bk)-\Go_{n}^{}$. Thanks to the first equalities in (\ref{dpG}) and (\ref{dpGn}) we have
\begin{equation*}
\Go(\bk)-\Go_{n}^{} =[\Go_{N}^{}(\bk)-\Go_{N,n}^{}]\overrightarrow{\Lo}_{N}\Go(\bk)+\Go_{N,n}^{}U'[\Go(\bk)-\Go_{n}^{}],
\end{equation*}
so that for $\bk$ in a vicinity of some $i\kappa_{n}$, $n=1,\dots,\No$, under the same assumptions as above, we derive from (\ref{GNa})
\begin{equation*}
\Go(\bk)-\Go_{n}^{} =a(\bk-i\kappa_{n})\Phi_{N,n}^{} \otimes\Psi_{N,n}^{}\overrightarrow{\Lo}_{N}^{}\Go(\bk)+\Go_{N,n}^{}U'[\Go(\bk)-\Go_{n}^{}]+o(1).
\end{equation*}
Thus again, as in derivation of (\ref{diffG}), we have by (\ref{dpGn}) that
\begin{equation}
\Go(\bk)-\Go_{n}^{} =a(\bk-i\kappa_{n})\Phi_{n}^{}
\otimes\Psi_{N,n}^{}\overrightarrow{\Lo}_{N}^{}\Go(\bk)+o(1),\label{GGn1}
\end{equation}
where notation (\ref{Phin1}) was used. In order to find the expression for $\Psi_{N,n}\overrightarrow{\Lo}_{N}\Go(\bk)$ we apply operation $\Psi_{N,n}\overrightarrow{\Lo}_{N}$ to the above equality from the left, that gives
\begin{equation*}
\Psi_{N,n}^{}\overrightarrow{\Lo}_{N}\Go(\bk)=\dfrac{\Psi_{n}^{}}{1-a(\bk-i\kappa_{n})c_n}+o(1),
\end{equation*}
recalling (\ref{Psin1}) and the second equality in (\ref{cn2}). Notice, that thanks to (\ref{ak1}) and (\ref{c4}) the denominator here is different from zero, so we get by (\ref{GGn1})
\begin{equation}
\Go(\bk)=\Go_{n}^{}+a(\bk-i\kappa_{n})\dfrac{\Phi_{n}^{}\otimes\Psi_{n}^{}}{1-c_na(\bk-i\kappa_{n})}+o(1)\label{GGn2} 
\end{equation}
for $\bk\sim{i}\kappa_{n}$ for any $n\in[1,\No]$.

\subsection{Behavior of $\Phi(\bk)$ and $\Psi(\bk)$ for $\bk\sim i\kappa_{n}$}

Thanks to (\ref{GGn2}) and definitions (\ref{2-6}) we can now derive the behavior of the Jost solutions in vicinities of the points of discontinuity. Thus, applying $\overleftarrow{\Lo}_{N}\Phi_{N}(\bk)$ from the right to (\ref{GGn2}) we get, for $\bk\sim{i}\kappa_{n}$
\begin{equation}
\Phi (x,\bk)=\dfrac{\Phi_{n}^{}(x)}{1-c_{n}^{}a(\bk-i\kappa_{n})}+o(1),\qquad\bk\sim{i}\kappa_{n},\quad n=1,\dots,\No, \label{PhiPhin}
\end{equation}
recalling that $\Phi_{N}(\bk)$ is continuous at $\bk=i\kappa_{n}$, and using equalities (\ref{Phin2}) and (\ref{cn}). Situation with $\Psi(x,\bk)$  is a bit more complicated due to the pole behavior of $\Psi_{N}(\bk)$ at these points. Thus, applying
$\Psi_{N}(\bk)\overrightarrow{\Lo}_{N}$ from the left to (\ref{GGn2}) we get, recalling (\ref{Psin2}) and (\ref{cn2})
\begin{equation}
\Psi (\bk)=\dfrac{i\Psi_{n}^{}}{(\bk-i\kappa_{n})(1-c_{n}^{}a(\bk-i\kappa_{n}))}+O(1),\qquad\bk\sim{i}\kappa_{n}.
\quad n=1,\dots,\No. \label{PsiPsin}
\end{equation}

We see that reductions of the relation (\ref{GGn2}) for the Green's functions describe the asymptotic behavior of the Jost solutions in vicinities of their discontinuities. Notice that this behavior is in agreement with (\ref{Phin2}) and (\ref{Psin2}) thanks to properties of the function $a(\bk)$ in (\ref{ak})--(\ref{ak2}). Additional relations on the discrete values of the Jost solutions follow by means of the reduction of (\ref{GmGn}). Say, applying $\overleftarrow{\Lo}_{N}\Phi_{N,n}$ to this equality from the right we get by (\ref{Phin1})
\begin{equation}
\Go_{m}^{}\overleftarrow{\Lo}^{}_{N}\Phi_{N,n}^{}=\Phi_{n}^{}+\sum_{l=m+1}^{n+\No}\dfrac{\Phi_{l}c_{ln}}{1-c_l},\label{Gm} 
\end{equation}
where we have introduced the constants
\begin{equation}
c_{mn}^{}=\bigl(\Psi_{m}^{}\overleftarrow{\Lo}_{N}^{}\Phi_{N,n}^{}\bigr),\quad m,n=1,\ldots,\No.\label{cmn}
\end{equation}
Definition (\ref{cmn}) thanks to (\ref{LLN}), (\ref{deJ}), and (\ref{Psin1}) can be written in the different forms
\begin{equation}
c_{mn}^{}=\bigl(\Psi_{N,m}^{}\overrightarrow{\Lo}_{N}^{}\Go_{m}^{}\overleftarrow{\Lo}_{N}^{}\Phi_{N,n}^{}\bigr)\equiv
\bigl(\Psi_{m}^{}U'\Phi_{N,n}^{}\bigr),\label{cmn2}
\end{equation}
where, say, the last equality is a short form for
\begin{equation}
c_{mn}^{}=\int dx\,u'(x)\Psi_{m}^{}(x)\Phi_{N,n}^{}(x).\label{cmn3}
\end{equation}
We see that the diagonal terms coincide with (\ref{cn}):
\begin{equation}
c_{nn}^{}=c_{n}^{}.\label{cnn}
\end{equation}
In contrast to the diagonal case, the off-diagonal constants looks to need more strict conditions on the perturbation $u'(x)$. Indeed, as follows from (\ref{Phinx}) the product $\Psi_{m}^{}(x)\Phi_{N,n}^{}(x)$ in (\ref{cmn3}) can grow exponentially, so $u'(x)$ must decay at space infinity faster than some exponent linear in $x$, in order that integrals in this equality converge.

Eq.~(\ref{Gm}) enables to establish relations between different discrete values, generalizing (\ref{Phid}). Thus, summing it up with the matrix $\Do_{nj}$ we cancel the term in the l.h.s.\ thanks to (\ref{Phid}), getting
\begin{equation}
\sum_{n=1}^{\No}\Phi_{n}^{}\Do_{nj}+\sum_{n=1}^{\No}\sum_{l=m+1}^{n+\No}
\dfrac{\Phi_{l}^{}c_{ln}^{}\Do_{nj}}{1-c_{l}^{}}=0.\label{PhiD1}
\end{equation}
The $m$-dependence of the second term is irrelevant: subtracting from (\ref{PhiD1}) the same equation with $m\to{m-1}$ we get, due to (\ref{Phid}), only a trivial consequence of (\ref{cmn})
\begin{equation}
\sum_{n=1}^{\No}c_{mn}^{}\Do_{nj}^{}=0. \label{cD}
\end{equation}
Let now $m=\No$ in (\ref{PhiD1}). This gives a generalization of the first equality in (\ref{Phid}) for the perturbed case:
\begin{equation}
(\Phi_{1}^{}(x),\ldots,\Phi_{\No}^{}(x))\t{\Do}=0, \label{PhiD2}
\end{equation}
where
\begin{align}
\t{\Do}_{nj}^{}&=\Do_{nj}^{}+\dfrac{1}{1-c_{n}^{}}\sum_{l=n}^{\No}c_{nl}^{}\Do_{lj}^{}\equiv
\Do_{nj}^{}-\dfrac{1}{1-c_{n}^{}}\sum_{l=1}^{n-1}c_{nl}^{}\Do_{lj}^{}\equiv\nonumber\\
&\equiv\dfrac{1}{1-c_{n}^{}}\Bigl(\Do_{nj}^{}+\sum_{l=n+1}^{\No}c_{nl}^{}\Do_{lj}^{}\Bigr). \label{tD}
\end{align}
the second and third equalities following from (\ref{cD}) and (\ref{cnn}), correspondingly. Since for the zero perturbation, $u'(x)\equiv 0$, the matrix $\t{\Do}$ reduces to the matrix $\Do$, that we choose to be TP, it is natural to assume that also the matrix $\t{\Do}$ is TP.

\section{Inverse problem}
\subsection{$\ol\partial$-derivatives of the Green's function and Jost solution}

In order to find the departure from analyticity of the Green's function we notice first that by (\ref{ak}) in the sense of distributions
\begin{equation}
\dfrac{\partial {a(\bk)}}{\partial {\ol{\bk}}}=\dfrac{\sgn{\bk_{\Re}^{}}}{2\pi {i}\ol{\bk}}.\label{ak3}
\end{equation}
Therefore, for the $\ol\partial$-bar derivative of the Green's function $\Go(\bk)$ in a neighborhood of a point
$\bk\sim{i}\kappa_{n}$ we have thanks to (\ref{GGn2}),
\begin{equation}
\dfrac{\partial\Go(\bk)}{\partial\ol{\bk}}=\dfrac{\sgn\bk_{\Re}^{}}{2\pi{i}(\ol\bk+i\kappa_{n})}\dfrac{\Phi_{n}^{}\otimes\Psi_{n}^{}} {[1-c_na(\bk-i\kappa_{n})]^{2}}+O(1),\label{ip1}
\end{equation}
that is also valid in the sense of distributions, where the expression in the r.h.s.\ is locally integrable. Then the derivative of (\ref{dpG}) sounds as
\begin{equation*}
\dfrac{\partial\Go(\bk)}{\partial\ol{\bk}}=\dfrac{\sgn\bk_{\Re}^{}}{2\pi}\,\Phi_{N}(-\ol{\bk})\otimes
\Psi_{N}(-\ol{\bk})\overrightarrow{\Lo}_{N}^{}\Go(\bk)+\Go_{N}^{}(\bk)U'\dfrac{\partial\Go(\bk)}{\partial\ol{\bk}},
\end{equation*}
where we used (\ref{dG}), (\ref{LLN}), and (\ref{deG}). Thanks to (\ref{realG}) and the second equality in (\ref{2-6}) this can be written as
\begin{equation}
\dfrac{\partial\Go(\bk)}{\partial\ol{\bk}}=\dfrac{\sgn\bk_{\Re}^{}}{2\pi}\,\Phi_{N}(-\ol{\bk})\otimes
\Psi(-\ol{\bk})+\Go_{N}^{}(\bk)U'\dfrac{\partial\Go(\bk)}{\partial\ol{\bk}}.\label{ip2} 
\end{equation}
In the r.h.s.\ of this equality we have a product of two distributions. Nevertheless, it is valid as thanks to (\ref{ip1}), (\ref{Phik}), (\ref{PsiPsin}), and (\ref{ak1}) for the coefficients of the singular terms in both sides we have
\begin{equation*}
\Phi_{n}^{}\otimes\Psi_{n}^{}=[1-c_na(\bk-i\kappa_{n})]\Phi_{N,n}^{}\otimes\Psi_{n}^{}+ 
\bigl[\Go_{N,n}^{}+a(\bk-i\kappa_{n})\Phi_{N,n}^{}\otimes\Psi_{N,n}^{}\bigr]U'\Phi_{n}^{}\otimes\Psi_{n}^{},
\end{equation*}
that is an identity, because for the first term in the second brackets we can use $\Go_{N,n}U'\Phi_{n}=\Phi_{n}-\Phi_{N,n}$, that in its turn follows from (\ref{dpGn}) and (\ref{Phin1}), and in the second term we use the second equality in (\ref{cn2}). Thus under the standard assumption of unique solvability of the integral equation (\ref{dpG}) and applying $\overleftarrow{\Lo}_{N}\Phi_{N}(-\ol{\bk})$ to it from the right we derive
\begin{equation*}
\dfrac{\partial\Go(\bk)}{\partial\ol{\bk}}=\dfrac{\sgn\bk_{\Re}^{}}{2\pi}\,\Go(\bk)\overleftarrow{\Lo}_{N}\Phi_{N}(-\ol{\bk})
\otimes\Psi (-\ol{\bk}),
\end{equation*}
that thanks to (\ref{realG}) and (\ref{2-6}) gives finally
\begin{equation}
\dfrac{\partial\Go(\bk)}{\partial\ol{\bk}}=\dfrac{\sgn\bk_{\Re}^{}}{2\pi}\,\Phi (-\ol{\bk})\otimes\Psi (-\ol{\bk}),\label{dbG}
\end{equation}
that due to the above consideration is valid in terms of distributions.

From (\ref{2-6}) and (\ref{dbG}), recalling that $\Phi_{N}(\mathbf{k)}$ is analytic, we get
\begin{equation}
\dfrac{\partial\Phi (x,\bk)}{\partial\ol{\bk}}=\Phi (x,-\ol{\bk})r(\bk), \label{dbPhi}
\end{equation}
with
\begin{equation}
r(\bk)=\dfrac{\sgn\bk_{\Re}^{}}{2\pi}\bigl(\Psi (-\ol{\bk})\overleftarrow{\Lo}_{N}\Phi_{N}^{}(\bk)\bigr), \label{ip3}
\end{equation}
which, by using (\ref{2-6}) and (\ref{realG}), can be rewritten as
\begin{equation}
r(\bk)=\dfrac{\sgn\bk_{\Re}^{}}{2\pi}\bigl(\Psi_{N}^{}(-\ol{\bk})\overrightarrow{\Lo}_{N}^{}\Go(\bk)\overleftarrow{\Lo}_{N}\Phi_{N}^{}(\bk)\bigr) .\label{ip4}
\end{equation}
From (\ref{ip3}), thanks to the behaviors of $\Psi (\bk)$ derived in (\ref{PsiPsin}), and thanks to (\ref{cmn}), (\ref{ak1}), we obtain that $r(\bk)$ is a singular function for all $\bk=i\kappa_{n}$, $n=1,\ldots,\No$ and has the following singular structure:
\begin{equation}
r(\bk)=\dfrac{c_{n}^{}\sgn\bk_{\Re}^{}}{2\pi{i}(\ol{\bk}+i\kappa_{n})(1-c_{n}^{}a(\bk-i\kappa_{n}))}+O(1), 
\quad\bk\sim i\kappa_{n},\quad n=1,\dots,\No.\label{ip5}
\end{equation}
Therefore, thanks to the behavior of $\Phi(\bk)$, see (\ref{PhiPhin}), we derive from (\ref{dbPhi}), , that
\begin{equation}
\dfrac{\partial\Phi(\bk)}{\partial\ol{\bk}}=\dfrac{\Phi_{n}^{}c_{n}^{}
\sgn\bk_{\Re}^{}}{2\pi{i}(\ol{\bk}+i\kappa_{n})[1-c_{n}^{}a(\bk-i\kappa_{n})]^{2}}+O(1),\quad 
\bk\sim i\kappa_{n}, \label{ip6}
\end{equation}
i.e., this derivative has integrable singularities in all points $\bk=i\kappa_{n}$.

\subsection{Integral equation of the Inverse problem}

Thanks to (\ref{ip6}) $\Phi (x,\bk)$ is smooth enough and its $\ol{\partial}$-derivative does not need any regularization.  But in order to derive the integral equation of the Inverse problem it is necessary to take the nonstandard normalization (\ref{Asymptk}) into account. Thanks to (\ref{dbPhi}) we have that
\begin{equation}
\Phi (x,\bk)e_{}^{i\bk{x}_1+\bk^{2}x_2}=X(x,\bk)-\dfrac{1}{\pi}\int\dfrac{d^{2}\bk'}{\bk'-\bk}
e_{}^{i\bk'{x}_1+{\bk'}^{2}x_2}r(\bk')\Phi (x,-\ol{\bk'}),\label{sip1}
\end{equation}
where $X(x,\bk)$ is an entire function of $\bk$. The integral term decays when $\bk\to\infty$, so by (\ref{Asymptk}) $X(x,\bk)$ is a polynomial of order $N_b$ with coefficient $i^{N_b}$ of the highest power. The new potential $\t{u}(x)$ is defined then in analogy to (\ref{asymptk3}) by
\begin{equation}
\t{u}(x) =-2\lim_{\bk\rightarrow\infty}(i\bk)^{-N_{b}+1}\partial_{x_{1}}
\bigl(e^{i\bk x_{1}+\bk^{2}x_{2}}\Phi(x,\bk)\bigr).\label{sip10}
\end{equation}
So, as follows from (\ref{sip1}),
\begin{equation}
\t{u}(x) =-2\lim_{\bk\rightarrow\infty}(i\bk)^{-N_{b}+1}X_{x_{1}}(x,\bk).\label{sip11}
\end{equation}
Thus in order to get a closed system of equations of the Inverse problem we need to determine  $X(x,\bk)$ in terms of the Jost solutions itself. In particular, values $X(x,i\kappa_n)$ must be expressed in terms of the values $\Phi_{n}(x)$ of the Jost solution by means of the asymptotic relations (\ref{PhiPhin}). Such expression will allow to use condition (\ref{PhiD2}). Because of the asymptotic behavior in (\ref{ip6}) the limit $\bk\sim{i}\kappa_{n}$ of the integral term in (\ref{sip1}) needs a special consideration. For this aim we use the following\cite{BPPP2002} definition of the principal value distribution:
\begin{equation}
\text{p.v.}\int\dfrac{d^{2}\bk'\,f(\bk')\sgn{\bk'_{\Re}}}{|\bk'|^{2}[1-c{a}(\bk')]^{2}} =
\lim_{\epsilon\rightarrow 0}\int\limits_{|\bk'|>\epsilon}\dfrac{d^{2}\bk'\,f(\bk')\sgn{\bk'_{\Re}}}{|\bk'|^{2}[1-c{a}(\bk')]^{2}},\label{ip8} 
\end{equation}
where $f(\bk)$ is an arbitrary test-function.

\noindent\textsl{\textbf{Statement.\/}} Let $\bk,\bk'\in\Cs$, $a(\bk)$ be defined in (\ref{ak}) and $c$ an arbitrary real constant. Then
\begin{enumerate}
\item the distribution (\ref{ip8}) can be given equivalently by any of the next two equalities
\begin{align}
\text{p.v.}\int\dfrac{d^{2}\bk'\,f(\bk')\sgn{\bk'_{\Re}}}{|\bk'|^{2}[1-c{a}(\bk')]^{2}} &=\dfrac{1}{2}\int d^{2}\bk'\,\sgn{\bk'_{\Re}}\dfrac{f(\bk')-f(-\ol{\bk'})}{|\bk'|^{2}[1-c{a}(\bk')]^{2}}\equiv\label{ip10} \\
&\equiv\int\dfrac{d^{2}\bk'\,[f(\bk')-\theta(1-|\bk'|)f(0)]\sgn{\bk'_{\Re}}}{|\bk'|^{2}[1-c{a}(\bk')]^{2}},\label{ip9} 
\end{align}
\item in terms of distributions the following asymptotic is valid:
\begin{align}
\dfrac{\sgn \bk'_{\Re}}{\overline{\bk'}(\bk'-\bk)[1-c{a}(\bk')]^{2}}&=
\text{p.v.}\dfrac{\sgn \bk'_{\Re}}{|\bk'|^{2}[1-c{a}(\bk')]^{2}}-\nonumber\\
&-\dfrac{2\pi^2{i}}{c}\Bigl(\dfrac{1}{1-c{a}(\bk)}+\dfrac{\log(1-c)}{c}\Bigr)\delta(\bk')+o(1),\quad\bk\sim0.\label{ip7} 
\end{align}
\end{enumerate}
\textsl{Proof.\/}  Equality (\ref{ip10})  results from (\ref{ak1}) thanks to $\sgn \bk'_{\Re}$ and does not need any regularization. This equality proves that definition (\ref{ip8}) is meaningful. Now (\ref{ip9}) follows from (\ref{ip10}) again thanks to (\ref{ak1}) and
$\sgn$-function. In order to prove (\ref{ip7}) we write
\begin{align*}
\int\dfrac{d^{2}\bk'\,f(\bk')\sgn \bk'_{\Re}}{\overline{\bk'}(\bk'-\bk)[1-ca(\bk')]^{2}}&=
\int\dfrac{d^{2}\bk'\,[f(\bk')-\theta(1-|\bk'|)f(0)]\sgn \bk'_{\Re}}{\overline{\bk'}(\bk'-\bk)[1-ca(\bk')]^{2}}+\\
&+f(0)\int\limits_{|\bk'|<1}\dfrac{d^{2}\bk'\sgn \bk'_{\Re}}{\overline{\bk'}(\bk'-\bk)[1-ca(\bk')]^{2}},
\end{align*}
where the first integral now admits limit $\bk\to0$. In the second term like in derivation of (\ref{ip1})  we assume that $c$ is different from zero and use that thanks to (\ref{ak3})
\begin{equation*}
\dfrac{\sgn{\bk'_{\Re}}}{\ol{\bk'}[1-c{a}(\bk)]^{2}}=\dfrac{2\pi{i}}{c}\ol{\partial^{}}_{\bk'}\dfrac{1}{1-ca(\bk')}.
\end{equation*}
Then
\begin{align*}
\int\dfrac{d^{2}\bk'\,f(\bk')\sgn \bk'_{\Re}}{\overline{\bk'}(\bk'-\bk)[1-ca(\bk')]^{2}}&=
\int\dfrac{d^{2}\bk'\,[f(\bk')-\theta(1-|\bk'|)f(0)]\sgn \bk'_{\Re}}{|\bk'|^{2}[1-ca(\bk')]^{2}}+\\
&+f(0)\dfrac{2\pi{i}}{c}\int\limits_{|\bk'|<1}d^{2}\bk'\dfrac{1}{\bk'-\bk}\dfrac{\partial}{\partial\overline{\bk'}}
\dfrac{1}{1-ca(\bk')}+o(1),\quad\bk\sim0.
\end{align*}
Thanks to (\ref{ip9})  the first term is the principal value distribution (\ref{ip8}). For the second term thanks to the Cauchy--Green formula for $|\bk|<1$ we have
\begin{equation*}
\int\limits_{|\bk'|<1}d^{2}\bk'\dfrac{1}{\bk'-\bk}\dfrac{\partial}{\partial\overline{\bk'}}
\dfrac{1}{1-ca(\bk')}=\dfrac{1}{2i}\oint\limits_{|\bk'|=1}d\bk'\dfrac{1}{(\bk'-\bk)[1-ca(\bk')]}-\dfrac{\pi}{1-ca(\bk)}.
\end{equation*}
Now we can perform the limit in the contour integral, getting
\begin{align*}
\lim_{\bk\to0}\oint\limits_{|\bk'|=1}d\bk'\dfrac{1}{(\bk'-\bk)[1-ca(\bk')]}&=\oint\limits_{|\bk'|=1}\dfrac{d\bk'}{\bk'[1-ca(\bk')]}=
2i\int\limits_{0}^{\pi}\dfrac{d\alpha}{1-c\alpha/\pi}=\\
&=\dfrac{-2\pi{i}}{c}\log(1-c),
\end{align*}
where (\ref{ak2}) was used. Summarizing, we have that
\begin{align*}
\int\dfrac{d^{2}\bk'\,f(\bk')\sgn \bk'_{\Re}}{\overline{\bk'}(\bk'-\bk)[1-ca(\bk')]^{2}}&=
\text{p.v.}\int\dfrac{d^{2}\bk'\,f(\bk')\sgn \bk'_{\Re}}{|\bk'|^{2}[1-ca(\bk')]^{2}}-\\
&-f(0)\dfrac{2\pi^{2}{i}}{c}\biggl[\dfrac{1}{1-ca(\bk)}+\dfrac{\log(1-c)}{c}\biggr]+o(1),\quad\bk\sim0.
\end{align*}
that proves (\ref{ip7}). The case $c=0$ follows then by the limiting procedure. $\blacksquare$

Now we can get values of $X(x,\bk)$ at points $\bk=i\kappa_{n}$, $n\in[1,\No]$, in terms of the discrete values (\ref{Phin1}) of the Jost solution. Thanks to (\ref{ip6}) the singular behavior of the integrand in (\ref{sip1}) at $\bk\sim{i}\kappa_{n}$ is exactly like in (\ref{ip7}),  once shifted $\bk$ and $\bk'$ for $-i\kappa_{n}$. We define
\begin{align}
&\text{p.v.}\int\dfrac{d^{2}\bk'}{\bk'-i\kappa_{n}^{}}e_{}^{i\bk'{x}_1+{\bk'}^{2}x_2}r(\bk')\Phi (x,-\ol{\bk'})=\nonumber\\
&\qquad=\lim_{\epsilon\to0}\int\limits_{|\bk'-i\kappa_{n}|>\epsilon}\dfrac{d^{2}\bk'}{\bk'-i\kappa_{n}^{}}
e_{}^{i\bk'{x}_1+{\bk'}^{2}x_2}r(\bk')\Phi (x,-\ol{\bk'}).\label{sip3}
\end{align}
so that by (\ref{sip1}) and (\ref{ip7})
\begin{align}
\Phi (x,\bk)e_{}^{i\bk{x}_1+\bk^{2}x_2}&=X(x,i\kappa_{n}^{})-\nonumber\\
&-\dfrac{1}{\pi}\text{p.v.}\int\dfrac{d^{2}\bk'}{\bk'-i\kappa_{n}^{}}e_{}^{i\bk'{x}_1+{\bk'}^{2}x_2}r(\bk')\Phi (x,-\ol{\bk'})+
\nonumber\\
&+\Phi_{n}(x)e_{}^{-K_{n}(x)}\biggl[\dfrac{1}{1-c_{n}^{}a(\bk-i\kappa_{n}^{})}+\dfrac{\log(1-c_{n}^{})}{c_{n}^{}}\biggr]+o(1),\quad\bk\sim{i}\kappa_{n},\label{ip11} 
\end{align}
where notation (\ref{Kn}) was used. Now by (\ref{PhiPhin}) we get that the undetermined terms on both sides cancel out and we arrive to
\begin{align}
X(x,i\kappa_{n}^{})&=\varphi_n(x)e_{}^{-K_{n}(x)}+\nonumber\\
&+\dfrac{1}{\pi}\text{p.v.}\int\dfrac{d^{2}\bk'}{\bk'-i\kappa_{n}}e^{i\bk'{x}_1+{\bk'}^{2}x_2}r(\bk')\Phi (x,-\ol{\bk'}),\label{ip12}
\end{align}
where we introduced
\begin{equation}
\varphi_n(x)=\dfrac{\log(1-c_{n}^{})}{-c_{n}^{}}\Phi_{n}(x),\quad, n=1,\ldots,\No.\label{ip13}
\end{equation}
Notice that the ratio $\dfrac{\log(1-c_{n})}{-c_{n}}$ is positive thanks to (\ref{c4}). Let in analogy to  (\ref{Phik})
\begin{equation}
\varphi(x)=\{\varphi_{1}(x),\ldots,\varphi_{\No}(x)\},\label{sip4}
\end{equation}
denotes a $\No$-row of rescaled solutions of the heat equation. Then (\ref{PhiD2}) is written as
\begin{equation}
\varphi(x)\t{\t\Do}=0,\label{sip5}
\end{equation}
where (see (\ref{tD}))
\begin{equation}
\t{\t\Do}_{nj}=\dfrac{-c_{n}^{}}{(1-c_{n}^{})\log (1-c_{n}^{})}
\biggl(\Do_{nj}^{}+\sum_{l=n+1}^{\No}c_{nl}^{}\Do_{lj}\biggr),\quad n=1,\ldots,\No,\quad j=1,\ldots,N_b,\label{sip7}
\end{equation}
that gives $N_b$ equations on $\No=N_a+N_b$ unknowns $\varphi_n(x)$.

\subsection{Closure of the Inverse problem}

Assuming that the values of the polynomial $X(x,\bk)$ at points $\bk=i\kappa_{n}$ are known, it can be written as
\begin{equation}
X(x,\bk)=\sum_{n=1}^{\No}X(x,i\kappa_{n}^{})\prod_{\substack{m=1 \\m\neq{n}}}^{\No}\dfrac{\kappa_{m}+i\bk}{\kappa_{m}-\kappa_{n}},\label{ip14}
\end{equation}
that, apparently, is a polynomial of power $\No-1$, in contradiction with the properties of this polynomial mentioned after (\ref{sip1}). In fact, in order the get an expression for $X(x,\bk)$, which explicitly displays the required properties, we need to take into account that the values $X(x,i\kappa_{n}^{})$ are not independent. In order to formulate these conditions, we denote by
\begin{equation*}
s_{l}^{}=s_{l}(\kappa_{1}^{},\ldots,\kappa_{\No}^{})=
\sum_{1\leq n_{1}^{}<\ldots <n_{l}^{}\leq\No}\kappa_{n_{1}}^{}\ldots\kappa_{n_{l}}^{},\qquad s_{0}^{}=1,
\end{equation*}
the symmetric polynomials of their arguments and by $s_{l}^{(n)}=s_{l}^{}(\kappa_{1}^{},\ldots,\h{\kappa_{n}^{}},\ldots\kappa_{\No}^{})$, $s_{0}^{(n)}=1$ the symmetric polynomials with omitted variable $\kappa_{n}$. Then
$s_{l}^{(n)}=\sum_{j=0}^{l}(-\kappa_{n}^{})_{}^{j}s_{l-j}^{}$ and
\begin{equation}
\prod_{\substack{m=1\\ m\neq{n}}}^{\No}(\kappa_{m}+i\bk)=\sum_{l=0}^{\No-1}(i\bk)_{}^{\No-1-l}s_{l}^{(n)}=
\sum_{l=0}^{\No-1}(i\bk)_{}^{\No-1-l}\sum_{j}^{l}(-\kappa_{n}^{})_{}^{j}s_{l-j}^{}, \label{ip15}
\end{equation}
Thus, (\ref{ip14}) thanks to (\ref{gamma}) takes the form
\begin{equation}
X(x,\bk)=(-1)^{\No-1}\sum_{l=0}^{\No-1}(i\bk)_{}^{\No-1-l}\sum_{j=0}^{l}s_{l-j}^{}
\sum_{n=1}^{\No}X(x,i\kappa_{n}^{})(-\kappa_{n}^{})_{}^{j}\gamma_{n}^{}. \label{ip16}
\end{equation}
In order to cancel extra powers all terms with $l=0,\ldots,N_{a}-2$ must be zero, where by (\ref{Nnanb}) $N_a=\No-N_b$. Choosing first $l=0$ we get that the term with $j=0$ in the sum over $j$ equals zero. Then both sums by $l$ and $j$ goes from 1. Now, in order to cancel the term with $l=1$ we have to put that term of the $j$-sum with $j=1$ equal to zero. Repeating this procedure up to $l=N_{a}-2$ we derive that all terms up to $j=N_{a}-2$ equal zero. Coefficient of the term with $l=N_{a}-1$ must be equal to $i^{N_b}$ in order to obey the normalization condition for polynomial $X(x,\bk)$  (see remark after (\ref{sip1})). All together this gives that the values of this polynomial must obey $N_a$ relations
\begin{equation}
\sum_{n=1}^{\No}X(x,i\kappa_{n}^{})\kappa_{n}^{l}\gamma_{n}=(-1)^{N_{b}}\delta_{l,N_{a}-1},\quad l=0,\ldots,N_{a}-1,\label{ip161}
\end{equation}
and representation (\ref{ip16}) takes the form
\begin{align}
X(x,\bk)&=\sum_{l=0}^{N_b}(i\bk)_{}^{N_b-l}s_{l}^{}+\nonumber\\
&+(-1)^{\No-1}\sum_{l=1}^{N_b}(i\bk)_{}^{N_b-l}
\sum_{j=1}^{l}s_{l-j}^{}\sum_{n=1}^{\No}X(x,i\kappa_{n}^{})(-\kappa_{n}^{})_{}^{j+N_a-1}\gamma_{n}^{}. \label{ip162}
\end{align}

Relations (\ref{ip161}) due to (\ref{ip12}) give
\begin{align}
\sum_{n=1}^{\No}\varphi_n(x)e_{}^{-K_{n}(x)}\kappa_{n}^{l}\gamma_{n}&+
\dfrac{1}{\pi}\text{p.v.}\int d^{2}\bk'\,e^{i\bk'{x}_1+{\bk'}^{2}x_2}r(\bk')\Phi (x,-\ol{\bk'})\sum_{n=1}^{\No}\dfrac{\kappa_{n}^{l}\gamma_{n}}{\bk'-i\kappa_{n}}=\nonumber\\
&=(-1)^{N_{b}}\delta_{l,N_{a}-1},\quad l=0,\ldots,N_a-1.\label{ip17}
\end{align}
Thanks to (\ref{gamma}) it is easy to check that
\begin{equation}
\sum_{n=1}^{\No}\dfrac{\kappa_{n}^{l}\gamma_{n}}{\bk'-i\kappa_{n}}=-i(-1)^{\No-l}_{}(i\bk')^{l}_{}\prod_{n=1}^{\No}\dfrac{1}{\kappa_{n}+i\bk'}, \quad l=0,\ldots,N_a,\label{ip18}
\end{equation}
so we can write relations (\ref{ip17}) in the form
\begin{equation}
\varphi(x){\gamma}e^{-K}\Vo^{\,\prime}=\bigl(\underbrace{0,\ldots,0,(-1)^{N_{b}}}_{N_{a}}\bigr)+v(x),\label{sip6}
\end{equation}
where we introduced $N_a$-row
\begin{align}
&v(x)=\{v_{1}^{}(x),\ldots,v^{}_{N_a}(x)\},\label{sip8}\\
&v_{l}(x)=\dfrac{(-1)^{\No-l}}{i\pi}\text{p.v.}\int d^{2}\bk'\dfrac{(i\bk')_{}^{l-1}e^{i\bk'x_1+{\bk'}^{2}x_2}r(\bk')\Phi (x,-\ol{\bk'})}{\displaystyle\prod_{n=1}^{\No}(\kappa_{n}+i\bk')},\label{sip9}
\end{align}
for $l=1,\ldots,N_a+1$, being the value $l=N_a+1$ needed below. Together with (\ref{sip5}) these equations give just $\No$ equations to determine $\No$ discrete values of the Jost solutions, that close the formulation of the inverse problem. In order to reconstruct the perturbed potential we notice that by (\ref{sip11})
\begin{equation*}
\t{u}(x) =-2\lim_{\bk\rightarrow\infty}(i\bk)^{-N_{b}+1}X_{x_{1}}(x,\bk),
\end{equation*}
so that since (\ref{ip162}) we get
 \begin{equation*}
\t{u}(x) =2(-1)^{N_b}_{}\partial_{x_1}^{}\sum_{n=1}^{\No}\kappa_{n}^{N_a}\gamma_{n}^{}X(x,i\kappa_{n}^{}),
\end{equation*}
that in its turn can be written thanks to (\ref{ip12}) and (\ref{ip18}) as
\begin{equation}
\t{u}(x) =2(-1)^{N_b}_{}\partial_{x_1}^{}\biggl(\sum_{n=1}^{\No}\kappa_{n}^{N_a}\gamma_{n}^{}e_{}^{-K_{n}(x)}\varphi_{n}^{}(x)-v_{N_a+1}^{}(x)\biggr),\label{sip13} \end{equation}
where $v_{N_a+1}(x)$ is defined in (\ref{sip9}).

Finally, we have to exclude $X(x,\bk)$ from the equation of the Inverse problem (\ref{sip1}). Inserting (\ref{ip12}) in (\ref{ip14}) we get
\begin{align*}
X(x,\bk)& =\sum_{n=1}^{\No}\varphi_{n}(x)e_{}^{-K_{n}(x)}
\prod_{\substack{m=1\\ m\neq n}}^{\No}\dfrac{\kappa_{m}+i\bk}{\kappa_{m}-\kappa_{n}}+\\
&+\dfrac{1}{\pi}\text{p.v.}\int d^{2}\bk'\,e_{}^{i\bk'{x}_1+{\bk'}^{2}x_2}r(\bk')\Phi (x,-\ol{\bk'})
\sum_{n=1}^{\No}\dfrac{1}{\bk'-i\kappa_{n}}\prod_{\substack{m=1\\ m\neq n}}^{\No}\dfrac{\kappa_{m}+i\bk}{\kappa_{m}-\kappa_{n}}.
\end{align*}
From the trivial identity
\begin{equation*}
\sum_{n=1}^{\No}\dfrac{1}{\bk'-i\kappa_{n}}\prod_{\substack{m=1\\ m\neq n}}^{\No}\dfrac{\kappa_{m}+i\bk}{\kappa_{m}-\kappa_{n}}=\dfrac{1}{\bk-\bk'}\Biggl[\prod_{n=1}^{\No}\dfrac{\kappa_{n}+i\bk}{\kappa_{n}+i\bk'}-1\Biggr]
\end{equation*}
we have that (\ref{sip1}) is reduced to the form
\begin{align}
\Phi (x,\bk)e_{}^{i\bk{x}_{1}+\bk^{2}x_2}& =\sum_{n=1}^{\No}\varphi_{n}(x)e_{}^{-K_{m}(x)}
\prod_{\substack{ m=1 \\ m\neq n}}^{\No}\dfrac{\kappa_{m}+i\bk}{\kappa_{m}-\kappa_{n}}+ \notag\\
&+\dfrac{1}{\pi}\text{p.v.}\int\dfrac{d^{2}\bk'}{\bk-\bk'}e_{}^{i\bk'x_1+{\bk'}^{2}x_2}r(\bk')\Phi (x,-\ol{\bk'})
\prod_{n=1}^{\No}\dfrac{\kappa_{n}+i\bk}{\kappa_{n}+i\bk'}. \label{ip19}
\end{align}
Thus the system of equations of the inverse problem is given by (\ref{ip19}), (\ref{sip5}), and (\ref{sip6}). In terms of solution of this system the potential is reconstructed by means of (\ref{sip13}).

\section{Time evolution}
In\cite{equivKPII} we proved that the multisoliton potential of the heat operator constructed by means of the Darboux transformation  coincides with the potential derived by the $\tau$-function approach. Thus we can use the standard scheme of this approach, see\cite{DJKM}, where the time evolution of the pure $N$ soliton solution of KPII is obtained by adding a time term to $K_{n}(x)$ defined in (\ref{Kn}), i.e., by writing
\begin{equation}
K_{n}(x,t)=\kappa_{n}x_{1}+\kappa_{n}^{2}x_{2}-4\kappa_{n}^{3}t.\label{227}
\end{equation}
Under this substitution $u_{N}(x)$ given in (\ref{ux}) obtaines time dependence and becomes the $N$ soliton solution of the KPII equation (\ref{KPII}). Correspondingly, the time evolution of the Jost solutions (\ref{PhiPsi}) is given then by the equations
\begin{align}
\partial_{t}\Phi_{N}(x,\bk)&=-\overrightarrow{\Ao_{N}}\Phi_{N}(x,\bk)+4i\bk^{3}\Phi_{N}(x,\bk),\label{238}\\ 
\partial_{t}\Psi_{N}(x,\bk)&=\Psi_{N}(x,\bk)\overleftarrow{\Ao_{N}}-4i\bk^{3}\Psi_{N}(x,\bk),\label{238:1}
\end{align}
where $\Ao_{N}(x,\partial)$ is given in (\ref{230}) with $u$ substituted by $u_{N}$ and where we used notation (\ref{JostandDual}) for the action of operator $\Ao(x,\partial)$ and its dual. Time evolution (\ref{227}) can be reformulated equivalently preserving definition of $K_{n}(x)$, but including time dependence in the spectral data of the $N$ soliton solution:
\begin{align}
&\dfrac{d\kappa_{n}}{dt}=0, \label{228}\\
&\dfrac{d\Do_{nm}}{dt}=-4\kappa_{n}^{3}\Do_{nm}+(\Do\alpha)_{nm},\qquad n=1,\dots\No,\quad m=1,\dots,N_{b},\label{229}
\end{align}
where $\alpha$ is an arbitrary (possibly time depending) $N_b\times{N}_{b}$ matrix corresponding to the remark that the $N$-soliton solution is fixed not by this matrix, but by a point on a Grassmanian. If we choose the matrix $\Do$ in some way, then the matrix $\alpha$ must be fixed correspondingly. Say, if we choose (without loss of generality) that $\Do_{nm}=\delta_{n,m}$ for all $m,n=1,\dots,N_{b}$ then $\alpha=\diag\{4\kappa_{1}^{3},\ldots,4\kappa_{N_b}^{3}\}$.

Let us notice that above results for the $N$ soliton solution are independent of the choice of the sign in the bottom limit of the integral term in (\ref{230}), as thanks to (\ref{ux}) and asymptotic behavior (\ref{tau:sn}) of the $\tau$-function, we have that $\int\limits^{+\infty}_{-\infty}dx_{1}\,\partial^{2}_{x_{2}}u_{N}(x_{1},x_{2})=0$. This situation is changed when we consider the perturbed solution $u(x)$. Understanding time dependence in all terms of (\ref{tu}) and taking that both $u(x)$ and $u_{N}(x)$ obey KPII equation (\ref{KPII}) into account, we get that perturbation $u'$ must obey
\begin{equation}
(u'_{t}-6u'u'_{x_{1}}-6(u'u^{}_{N})_{x_{1}}+u'_{x_{1}x_{1}x_{1}})_{x_{1}}=-3u'_{x_{2}x_{2}}.\label{KPII'}
\end{equation}
Again, using that $u_{N}(x)$ exponentialy decays when $x_1\to\pm\infty$, we see that asymptotically $u'$ is governed by the KPII equation, more exactly by its linear part. Thus according to an analysis analogous to that performed in \cite{KPI94} for the KPI equation, the evolution form of (\ref{KPII'}) is given by
\begin{equation}
u'_{t}-6u'u'_{x_{1}}-6(u'u^{}_{N})_{x_{1}}+u'_{x_{1}x_{1}x_{1}}=-3\int\limits_{t\infty}^{x_1}dx'_{1}u'_{x_{2}x_{2}}(x'_{1},x^{}_{2}),\label{KPII''}
\end{equation}
where $t\infty$ denotes sign of the infinity limit. Thanks to the above discussion, the perturbed solution $u(x)$ must satisfy the evolution equation
\begin{equation}
u_{t}(x)-6u(x)u_{x_{1}}(x)+u_{x_{1}x_{1}x_{1}}(x)=-3\int_{t\infty}^{x_{1}}dx_{1}'u_{x_{2}x_{2}}(x_{1}',x_{2}),
\label{230.2}
\end{equation}
so that for generic intial data the time derivative of the solution is discontinuouse at $t=0$. In order to get this form of the evolution equation by means of the compatibility condition (\ref{LaxN}) we have to modify, correspondingly, the definition of operator $\Ao$ in (\ref{230}) correspondingly:
\begin{equation}
\Ao(x)=4\partial_{x_{1}}^{3}-6u\partial_{x_{1}}-3u_{x_{1}}-3\int\limits_{t\infty}^{x_{1}}dx'_{1}\,
u_{x_{2}}(x_{1}',x_{2}).\label{230:1}
\end{equation}

In order to find out the time evolution of the Jost solution $\Phi(x,\bk)$ we notice that thanks to (\ref{Phik}), (\ref{Psik}) and (\ref{238}), (\ref{238:1})
\begin{align}
\partial_{t}\Phi_{N,n}(x)&=-\overrightarrow{\Ao_{N}}\Phi_{N,n}(x)+4\kappa^{3}_{n}\Phi_{N,n}(x),\label{239:02}\\ 
\partial_{t}\Psi_{N,n}(x)&=\Psi_{N,n}(x)\overleftarrow{\Ao_{N}}-4\kappa^{3}_{n}\Psi_{N,n}(x),\label{239:03}
\end{align}
$n=1,\ldots,\No$, so that by (\ref{g3})
\begin{equation}
\partial_{t}\Go_{N}=[\Go_{N},\Ao_{N}].\label{239}
\end{equation}
Now differentiating (\ref{tphi}) by $t$ we use that (\ref{KPII''}) can be written thanks to (\ref{LaxN}) and (\ref{tu}) as
\begin{equation}
u'_{t}=[\Lo_{N},\Ao_{N}]-[\Lo,\Ao],\label{239.1}
\end{equation}
that by means of (\ref{238}) and (\ref{239}) enables derivation of the standard relations
\begin{equation}
\partial_{t}\Phi(x,\bk)=-\overrightarrow{\Ao}\Phi(x,\bk)+4i\bk^{3}\Phi(x,\bk),\qquad
\partial_{t}\Psi(x,\bk)=\Psi(x,\bk)\overleftarrow{\Ao}-4i\bk^{3}\Psi(x,\bk). \label{240}
\end{equation}

Then  derivative of (\ref{dbPhi}) by $t$ gives
\begin{equation}
\partial_{t}r(\bk,t)=4i(\bk^{3}+\ol{\bk}^{3})r(\bk,t), \label{244}
\end{equation}
so that from the singular behavior of $r(\bk)$, $\Phi(x,\bk)$ and $\Psi(x,\bk)$ (see (\ref{ip5}) and (\ref{PhiPhin}), (\ref{PsiPsin})) we derive
\begin{align}
&\dfrac{dc_{n}^{}}{dt}=0,\label{246}\\
&\partial_{t}\Phi_{n}(x)=-\overrightarrow{\Ao}\Phi_{n}(x)+4\kappa^{3}_{n}\Phi_{n}(x),\quad\text{for }n=1,\dots,\No, \label{242}\\
&\partial_{t}\Psi_{n}(x)=\Psi_{n}(x)\overleftarrow{\Ao}-4\kappa^{3}_{n}\Psi_{n}(x).\label{242:1}
\end{align}
Thus by (\ref{ip13})
\begin{equation}
\partial_{t}\varphi_{n}(x)=-\overrightarrow{\Ao}\varphi_{n}(x)+4\kappa^{3}_{n}\varphi_{n}(x),\quad\text{for }n=1,\dots,\No, \label{242:2}
\end{equation}
and following the same procedure as in derivation of (\ref{240}) we get by (\ref{239.1}), (\ref{239:02}) and (\ref{242:1}) for the time derivative of constants $c_{mn}$ defined in (\ref{cmn2})
\begin{equation}
\dfrac{dc_{mn}^{}}{dt}=4(\kappa_{n}^{3}-\kappa_{m}^{3})c_{nm}^{},\qquad n,m=1,\dots,\No. \label{251}
\end{equation}
Now from (\ref{sip7}), thanks to (\ref{229}), (\ref{246}) and (\ref{251}), we derive
\begin{equation}
\dfrac{d\t{\t\Do}_{nm}}{dt}=-4\kappa_{n}^{3}\t{\t\Do}_{nm}+(\t{\t\Do}\alpha)_{nm},\qquad n=1,\dots\No,\quad m=1,\dots,N_{b}.
\end{equation}

\section{Conclusion}
We developed here a modification of the IST that enables to consider solutions of KPII of the kind (\ref{tu}). The study of a heat operator (\ref{heatop}) with such potentials, having ray behavior on the $x$-plane, cannot be considered, strictly speaking, a ``scattering problem''. Nevertheless, in terms of this modification we succeeded in introducing Jost solutions and scattering data and we proved that the mentioned asymptotic behavior of the potential results in specific discontinuities and singularities of these objects with respect to the spectral parameter $\bk$. Consequently, we gave a corresponding modification of the Inverse problem and time evolution of the scattering data. The proof of the unique solvability of the Direct (\ref{tphi}) and Inverse problem (\ref{sip5}), (\ref{sip6}), and  (\ref{ip19}) was not our aim here. The whole construction is based on the assumption of the unique solvability of (\ref{tphi}) for the class of potentials under consideration. While the corresponding proof must be a rather straight generalization of the case of a decaying potential (see\cite{BarYaacov}--\cite{Grinevich0}), since we have boundedness property (\ref{Green}), the investigation of the Inverse problem is much more involved and may need more details on properties of the scattering data. Here we mention only that  (\ref{sip5}) and (\ref{sip6}) give a system of linear equations on the $\No$-row $\varphi(x)$ in (\ref{sip4}). Relation (\ref{sip5}) can be considered a generalization to the perturbed case of relation (\ref{Phid}) of the pure soliton case. If, in analogy to (\ref{dd}), there exists  a matrix $\t{\t\Do}{^{\,\prime}}$ (with at least one nonzero maximal minor) such that $\t{\t\Do}{^{\,\prime}}\t{\t\Do}=0$, then (\ref{sip5}) means that there exists an $N_a$ row $\t\varphi(x)$ such that
\begin{equation}
\varphi(x)=\t\varphi(x){\t{\t\Do}}{^{\,\prime}}.\label{c2}
\end{equation}
Then, for the unique solvability of relation (\ref{sip6}) it is necessary that the determinant
\begin{equation}
\t\tau{^{\,\prime}}(x)=\det{\t{\t\Do}}{^{\,\prime}}\gamma{e}^{-K}\Vo^{\,\prime}\label{c3} \end{equation}
is different from zero. Relation (\ref{c3}) is analogous to the second representation for the pure soliton $\tau$-function in (\ref{tau}). Correspondingly, we inherit from the pure soliton case the condition that the matrix $\t{\t\Do}$ is TNN, or TP. Taking (\ref{c4}) into account, we get by (\ref{sip7}) that the matrix $\Do_{nj}+\sum_{l=n+1}^{\No}c_{nl}\Do_{lj}$, where the matrix $c_{nl}$ is constrained by (\ref{cD}), must have the same property. It is also necessary to mention that the diagonal elements of the matrix $c$ are related to the singular behavior of the spectral data $r(\bk)$, see (\ref{cnn}) and (\ref{ip5}). Say, if $r(\bk)\equiv0$ we have that all diagonal elements are zero. In this case the Inverse problem (\ref{sip5}), (\ref{sip6}), and  (\ref{ip19}) supply us with another $N$ soliton solution. This solution is given by a $\tau$-function (\ref{c3}) parameterized by the matrix  $\Do_{nj}+\sum_{l=n+1}^{\No}c_{nl}\Do_{lj}$ instead of the matrix $\Do_{nj}$. A formulation of conditions for this matrix to be TP, or at least TNN, as well for the above mentioned unique solvability, in the generic case, of the Direct and Inverse problems are open at this moment.

\section*{Acknowledgements}

This work is supported in part by the grants RFBR \# 11-01-00440 and \# 11-01-12037, Scientific Schools 4612.2012.1, by the Program of RAS ``Mathematical Methods of the Nonlinear Dynamics,'' by INFN, by MIUR (grant PRIN 2008 ``Geometrical methods in the theory of
nonlinear integrable systems''), and by Consortium E.I.N.S.T.E.IN. M.B.\ and F.P.\ acknowledge hospitality at the Department of
Mathematics and Physics of the University of Salento, Italy.

\end{document}